\documentclass[11pt]{article}
\usepackage[utf8]{inputenc}
\usepackage{caption}
\usepackage{braket,slashed,bm}
\usepackage{jheppub}
\usepackage{simplewick}
\usepackage{array,multirow}
\usepackage{amsmath} \DeclareMathAlphabet{\mathpzc}{OT1}{pzc}{m}{it}
\usepackage[normalem]{ulem}
\usepackage{calligra}
\usepackage{floatrow}
\DeclareMathAlphabet{\mathpzc}{OT1}{pzc}{m}{it}
\usepackage{mathrsfs}
\usepackage{tikz}

\usepackage{rotating}

\usepackage{pgf}

\usepackage{subcaption}
\usepackage{lscape}
\usepackage{hyperref}
\usepackage{graphicx}
\usepackage{booktabs}
\newfloatcommand{capbtabbox}{table}[][\FBwidth]

\graphicspath{{./Figures/}}

\def\beq{\begin{equation}}
\def\eeq{\end{equation}}
\def\bea{\begin{eqnarray}}
\def\eea{\end{eqnarray}}
\def\nn{\nonumber \\}

\renewcommand{\to}{\rightarrow}

\pgfkeys{/pgf/number format/.cd ,precision=2,sci generic={exponent={\times 10^{#1}}}}

\allowdisplaybreaks

\begin{document}

\title{Monolepton production in SMEFT to $\mathcal O(1/\Lambda^4)$ and beyond}

\author[a]{Taegyun Kim}
\emailAdd{tkim12@nd.edu}
\author[a]{and Adam Martin.}
\emailAdd{amarti41@nd.edu}
\affiliation[a]{Department of Physics, University of Notre Dame, Notre Dame, IN, 46556, USA}
\abstract{We calculate $pp \to \ell^{+}\nu, \ell^-\bar \nu$ to ${\cal{O}}(1/\Lambda^4)$ within the Standard Model Effective Field Theory (SMEFT) framework. In particular, we calculate the four-fermion contribution from dimension six and eight operators, which dominates at large center of mass energy. We explore the relative size of the $\mathcal O(1/\Lambda^4)$ and $\mathcal O(1/\Lambda^2)$ results for various kinematic regimes and assumptions about the Wilson coefficients. Results for Drell-Yan production $pp \to \ell^+\ell^-$ at ${\cal{O}}(1/\Lambda^4)$ are also provided. Additionally, we develop the form for four fermion contact term contributions to $pp \to \ell^{+}\nu, \ell^-\bar \nu, pp \to \ell^+\ell^-$ of arbitrary mass dimension. This allows us to estimate the effects from even higher dimensional (dimension $> 8$) terms in the SMEFT framework.}

\maketitle

\section{Introduction}
\label{sec:intro}

 In this paper we calculate monolepton production $pp \to \ell^+ \nu, \ell^- \bar{\nu}$  within the Standard Model effective field theory (SMEFT)~\cite{Buchmuller:1985jz, Grzadkowski:2010es,Brivio:2017vri} out to $\mathcal O(1/\Lambda^4)$. The framework for SMEFT consists of the SM supplemented by a series of higher dimensional operators formed from the SM fields and their (covariant) derivatives,
 \begin{align}
 \label{eq:SMEFT}
 \mathcal L_{SMEFT} = \mathcal L_{SM} + \sum\limits_{d=5}\sum\limits_{i = 1}^{n} \frac{C^{(d)}_i\,\mathcal O^{(d)}_i(Q,u_c,d_c,L,e_c,H, B_{\mu\nu},W^I_{\mu\nu}, G^{A}_{\mu\nu}; D_\mu)}{\Lambda^{d-4}}
 \end{align}
 where $d$ labels the mass dimension, $i$ runs over the set of independent operators at a given dimension, and $\Lambda$ is the new physics scale. By convention, we will use a single scale $\Lambda$ and absorb any differences in the scale or perturbation order when an operator is generated into the coefficients. Any UV physics with all states heavier than the SM and single source of electroweak symmetry breaking can be mapped to $\mathcal L_{SMEFT}$, though in this paper we will use it from the bottom up, meaning we will keep all effects to a given order in $1/\Lambda$ and refrain from imposing relations among different Wilson coefficients (the $C^{(d)}_i$). We will also assume baryon and lepton number are preserved. This eliminates all odd $d$ from Eq.~\eqref{eq:SMEFT} -- so that the sum starts at $d = 6$ -- and reduces the number of operators at even $d$.
 
An important feature of the SMEFT is that certain $d > 4$ operators can do more than create new vertices. Setting the Higgs to its vacuum expectation value (vev), it is possible for higher dimensional terms to feed back into lower dimensional quantities, altering couplings and the normalization of fields by terms proportional to  ${\cal{O}}((2\langle H^\dagger H \rangle)^n/\Lambda^{2n})$. Thus, there are three ways that SMEFT effects can enter into monolepton production:
\begin{itemize}
\item Higher dimensional operators impact how parameters of the electroweak SM Lagrangian are related to experimental data~\cite{Grinstein:1991cd, Alonso:2013hga, Hays:2018zze}. All corrections of this type scale as ${\cal{O}}((2\langle H^\dagger H \rangle)^n/\Lambda^{2n})$ $\equiv \frac{\bar v^{2n}_T}{\Lambda^{2n}}$, where $v_T$ is the minimum of the full Higgs potential and $n$ is an integer related to the mass dimension of the SMEFT operator\footnote{By full potential, we mean the potential including  SMEFT effects. Therefore, $\bar v_T$ is different from the Lagrangian parameter $v_0$. The two can be related, but this is often unnecessary, as $\bar v_T$ is what appears in the $W/Z$ masses and couplings and is therefore what is connected to $G_F$. The distinction between $\bar v_T$ and $v_0$ is only needed when multi-Higgs operators, such as $|H|^6$, etc. explicitly play a role in the process.}. The ratio of the Higgs vev squared to the new physics scale squared is our main expansion parameter, which we will call $x = \bar v^2_T/\Lambda^2$ for simplicity. We are interested in $\mathcal O(1/\Lambda^4)$ effects, so $\mathcal O(x^2)$. We will use the terms $\mathcal O(x,x^2) \leftrightarrow \mathcal O(1/\Lambda^2, 1/\Lambda^4)$ interchangeably throughout. 
\item SMEFT operators alter the $\bar f f W$ interactions of SM fermions with $W$ and $Z$ bosons. Effects of this type also scale as $\frac{\bar v^{2n}_T}{\Lambda^{2n}} \equiv x^n$.
\item Contact, four fermion $+X$ operators, where $X$ is some combination of Higgs bosons and/or derivatives. This type of effect may scale as $\frac{\bar v^{2n}_T}{\Lambda^{2n}},\frac{\hat s^{2n}}{\Lambda^{2n}}$ or $\frac{\bar v^{n}_T\,\hat s^{n/2}}{\Lambda^{2n}}$, where $\hat s$ is the energy (squared) of the process;  these scalings can rewritten as $x^n, \hat s^n \frac{x^n}{\bar v^{2n}_T}$ and $\hat s^{n/2}\frac{x^n}{\bar v^n_T}$.
\end{itemize}
The first two types of SMEFT correction can be arranged into compact, all-orders expressions via the geoSMEFT basis~\cite{Helset:2020yio}, and the electroweak parameter shifts and $\bar f f W$ corrections to $\mathcal O(1/\Lambda^4)$ were calculated in Ref.~\cite{Hays:2020scx, Corbett:2021eux}. As we will review in Sec.~\ref{sec:thesetup}, the fact that all-orders forms are possible for the first two SMEFT effects is related to the special kinematics of 2- and 3-particle vertices. Contact terms, meanwhile, involve 4 particles (at least), and therefore cannot be massaged into all-orders forms -- so the number and form of the vertices must be worked out manually at each order. 

Extending out $\mathcal O(1/\Lambda^4)$ will allow us to explore monolepton production within SMEFT in a wider range of energies. Purely by dimensional analysis, contact terms grow with energy faster the vertex or input corrections. Therefore, at $\hat s > v$, the largest SMEFT effects will be driven by dimension six contact terms that scale as (in the amplitude) ($\hat s/\Lambda^2)$. Understanding the exceptions to this naive scaling, and how it is impacted by truncation errors, requires knowing the next order term in the EFT expansion, $\mathcal O(1/\Lambda^4)$. This analysis is part of a larger program to better understand and quantify the uncertainty in SMEFT analyses coming from terms of higher mass dimension, the `truncation uncertainty'. Processes where the complete $\mathcal O(1/\Lambda^4)$ terms are known can be used as test cases and studied to derive estimates for scenarios where $\mathcal O(1/\Lambda^4)$ results are not known. 
As the state of the art for Monte Carlo SMEFT programs contain dimension six operators, the extent to which dimension six squared pieces can be used as a proxy for the complete $\mathcal O(1/\Lambda^4)$ result is particularly interesting.

Results for neutral current cousin $pp \to  \ell^+_j \ell^-_j$ within SMEFT can be obtained using the same steps. As the $\mathcal O(1/\Lambda^4)$ corrections to this process have been presented previously in Ref.~\cite{2020,Boughezal:2021tih}, our main focus will be monolepton production rather than dileptons. However, we present results for the dilepton case as well (though only at lowest order in $\alpha_s$),  for the sake of completeness, to present more analytic form for how the Wilson coefficients enter, and because our basis -- a geoSMEFT compatible basis -- differs slightly from what was used in Ref.~\cite{2020,Boughezal:2021tih}.

The organization of this paper is as follows. In Sec.~\ref{sec:thesetup} we introduce the operators at dimension six and eight that can play a role in monolepton and dilepton production and classify them according to whether they have a direct (meaning new vertices) or indirect effect. When converting the contact terms to Feynman rules, we  pay particular attention to their flavor and momentum structures. Section~\ref{sec:Wpparton} contains some parton level results. We focus on the large $\hat s$ pieces of the calculation, as the terms growing with energy have the most significant impact on the total cross section. Folding in parton distribution functions, we present proton level results in Sec.~\ref{sec:results}. We explore the impact of the $\mathcal O(1/\Lambda^4)$ terms, including what fraction comes from dimension eight operators versus dimension six squared. We also explore how our results change as we vary the relative sign and size of the Wilson coefficients. Then, in Sec.~\ref{sec:allorders} we look beyond dimension eight, deriving a compact formula (and count) for four fermion operators with arbitrary powers of Higgses and derivatives; this allows us to estimate the effects of $\mathcal O(1/\Lambda^6)$ SMEFT terms on monolepton production. Section~\ref{sec:conclude} contains our conclusions. The main text is supplemented by several appendices which contain results for dilepton production (analytic and plots analogous to Sec.~\ref{sec:results}) and results for alternate electroweak input schemes.

\section{Setup}
\label{sec:thesetup}

In the limit of massless SM fermions, the partonic amplitude can be classified by the helicity of the quarks $\lambda_f$  (as they must be the same within the SM) and the helicity of the outgoing leptons $\lambda_\ell$; $\lambda = L,R$. Each helicity class can be decomposed into a spinor piece $S_{\lambda_f, \lambda_\ell}$ and a coefficient $\mathcal A_{\lambda_f,\lambda_\ell}$. For the charged current, $A_{\lambda_f,\lambda_\ell}$ consists of the $W$ exchange piece plus any contact terms $a_{\lambda_f, \lambda_\ell}$
\begin{align}
\label{eq:incc}
\mathcal A^{CC, total}_{\lambda_f, \lambda_{\ell}} S_{\lambda_f,\lambda_\ell} =\Big(  \mathcal A^{W}_{\lambda_f, \lambda_{\ell}} (\hat s; m_W, \Gamma_W, g^{eff}_{W, f_L},g^{eff}_{W, f_R}, V_{CKM}) + \mathcal A^{4f}_{\lambda_f, \lambda_{\ell}}(a^{eff}_{\lambda_f, \lambda_{\ell}}) \Big)\, S_{\lambda_f,\lambda_\ell}.	
\end{align}
The neutral current contains a photon exchange piece, a $Z$ exchange piece, and contact terms: 
\begin{align}
\mathcal A^{NC, total}_{\lambda_f, \lambda_{\ell}} S_{\lambda_f,\lambda_\ell} =\Big( \mathcal A^{\gamma}_{\lambda_f, \lambda_{\ell}} (\hat s; e_{em}) + \mathcal A^{Z}_{\lambda_f, \lambda_{\ell}} (\hat s; m_Z, \Gamma_Z, g^{eff}_{Z, f_L},g^{eff}_{Z, f_R}) + \mathcal A^{4f}_{\lambda_f, \lambda_{\ell}}(a^{eff}_{\lambda_f, \lambda_{\ell}}) \Big)\, S_{\lambda_f,\lambda_\ell}	,
\label{eq:helamp}
\end{align}
where we have listed the parameters in each piece of the amplitude. Clearly, the contact terms come from higher dimensional operators. However, as mentioned earlier SMEFT effects also leak into the masses and three point couplings of SM fields.

Our first step is to list the operators at dimension six and eight that can play a role in these processes and to classify their effects. Before proceeding, let us define our flavor assumptions. Our baseline flavor assumption is $(U(3))^5 = U(3)_Q\otimes U(3)_{u_c} \otimes U(3)_{d_c} \otimes U(3)_L \otimes U(3)_{e_c}$ and CP invariance. This implies unit CKM and PMNS matrices\footnote{Provided we limit the final state leptons to electrons and muons, our flavor assumption can be relaxed actually $(U(2))^5$ without any change in our results, given the $b/\bar b/t/\bar t$ parton distributions are significantly smaller than the first two generation quarks. However, for simplicity, we will use $U(3)$ when discussing flavor assumptions.}. We make this assumption because we are primarily interested in the high energy behavior of dimension six and eight operators in $pp \to \ell^{\pm}\nu$, as this is where questions of higher dimensional effects can have exaggerated effects (relative to lower orders) and questions of EFT validity arise. The $(U(3))^5$ assumption simplifies the calculation of the large $\hat s$ regime while still capturing the physics we care about. For most purposes, the flavor symmetry can be relaxed to $U(3)_Q \times U(3)_{u+d} \times U(3)_{L}\times U(3)_{e_c}$, where $U(3)_{u + d}$ means a transformation where the right handed up and down quarks transform identically. We will comment on the differences between this looser flavor assumption and full $(U(3))^5$ where present. The ramifications of further loosening the flavor assumption are interesting and will be discussed for each operator category. The strategy of presenting a baseline flavor symmetry then successively loosening it is inspired by the approach in Ref.~\cite{saavedra2018interpreting}.

\subsection{Operators at dimension six}

The operators at dimension six can be grouped into those that affect field normalizations and couplings, those that affect trilinear vertices involving two fermions and a gauge field, and contact terms. For all operators, we exclusively work with left handed fermion fields.

The first class contains:
\begin{align}
\label{eq:d6type1}
& C^{(6)}_{HB}H^{\dag}H B^{\mu\nu}B_{\mu\nu} + C^{(6)}_{HW}H^{\dag}H W^{I,\mu\nu}W^I_{\mu\nu} + C^{(6)}_{HWB} H^{\dag}\tau^I H\, W^{\mu\nu}_I\, B_{\mu\nu}\nn
& \quad\quad\quad + C_{H\Box} (H^{\dag}H)\Box(H^{\dag}H) + C_{H D}^{(6)}\left(H^{\dagger} D_{\mu} H\right)^{\star}\left(H^{\dagger} D^{\mu} H\right).
\end{align}
and is clearly independent of our flavor assumption. Here we have left off the factors of $\Lambda$ for brevity.

The operators in the second class are
\begin{align}
\label{eq:d6type2}
& C^{1,(6)}_{H\psi} i (H^{\dag}\overleftrightarrow D_\mu H)\, (\psi^{\dag}_i \bar{\sigma}^\mu \psi_i),\quad \psi \in \{Q,u_c, d_c, L, e_c\} \nn
& C^{3,(6)}_{H\psi} i (H^{\dag}\overleftrightarrow D^I_\mu H)\, (\psi^{\dag}_i \bar{\sigma}^\mu\tau_I \psi_i),\quad \psi \in \{Q, L\}, \nn
& \textcolor{blue}{C^{(6)}_{Hud}\, i (\tilde H^\dag D_\mu H) (d^{\dag}_{c,i} \bar\sigma^{\mu} u_{c,i}) + h.c.}
\end{align}
Here $i$ is a flavor index, $\tilde H = \epsilon (H)^*$, $H^{\dag}\overleftrightarrow D_\mu H = H^{\dag}D_\mu H - (D_\mu H^{\dag})H$ and $H^{\dag}\overleftrightarrow D^I_\mu H = H^{\dag}\tau^I D_\mu H - (D_\mu H^{\dag})\tau^I\,H$ with $\tau^I$ the Pauli matrices. The $ C^{(6)}_{Hud}$ operators, which create right handed quark currents when the Higgs is set to its vev, are allowed by $U(3)_Q\times U(3)_{u+d}\times U(3)_{L}\times U(3)_{e_c}$ but are forbidden by $(U(3))^5$, hence we have colored them in blue. Reducing the flavor symmetry further rapidly increases the number of dimension six effects, as dimension six operators do not need to interfere with the SM to contribute to $\mathcal O(1/\Lambda^4)$ and are therefore insensitive to its helicity and flavor structure. For example, reducing the symmetry to  $U(3)_{Q+u+d} \times U(3)_{L+e}$, dipole terms are allowed. These have the wrong helicity structure to interfere with the SM, but appear unsuppressed when squared at $\mathcal O(1/\Lambda^4)$. 

The operators we've written above are in the Warsaw basis~\cite{Grzadkowski:2010es}. While useful for its familiarity, we also use it because it complies with a more general organization scheme known as geoSMEFT~\cite{Helset:2020yio}. A simple way to state the philosophy of geoSMEFT is to minimize the number of operators that contribute to two and three point vertices via judicious choices of where to put derivatives. Sticking with this philosophy and considering higher and higher dimensional operators, the number of operators that can contribute to a given two or three point vertex plateaus to a small, constant number (at each order). As such, it is possible to derive results to all orders. The plateau in the operator count is a consequence of trivial kinematics  -- meaning all dot products of momenta can be reduced to masses -- and a limited set of electroweak contractions one can make using two to three SM fields. The minimum set of two and three point vertices was established in Ref.~\cite{Helset:2020yio} in the form of `metrics' or `connections' on the 4-dimensional manifold corresponding to $\phi_I, I = 1\cdots4$, the four real degrees of freedom in the Higgs doublet. The operators in Eq.~\eqref{eq:d6type1}, \eqref{eq:d6type2} lie in the following metrics:
\begin{align}
h_{IJ}(\phi)(D_\mu\phi)^I (D^\mu\phi)^J, \quad g_{AB}(\phi)\mathcal W^{A,\mu\nu}\mathcal W^B_{\mu\nu}, \quad L^{\psi}_{I,A}(\phi)\,((D_\mu\phi)^I\,(\psi^{\dag}\bar\sigma^\mu \tau^A \psi)),
\end{align}
where $A$ runs from 1 to 4 and all four gauge bosons are grouped into $\mathcal W^A_{\mu\nu} = \{W^{1\cdots 3}_{\mu\nu}, B_{\mu\nu} \}$. The couplings and masses of the SM Lagrangian can be expressed compactly in terms of these metrics. For example, the terms relevant for monolepton production are\footnote{With our baseline flavor assumption and massless fermions, the CKM remains a unit matrix to all orders in $x$. Dropping these restrictions, the flavor information or a SMEFT theory can also be expressed in terms of the metrics, see Ref.~\cite{Talbert:2021iqn}.}: 
\begin{align}
\label{eq:geosmeft}
& g^{eff}_{W, f_L} = -\frac{\bar{g}_{2}}{\sqrt{2}} ( 1 - \bar v_T \langle L_{1,1}^{\psi, p r}\rangle \pm i \bar{v}_{T}\langle L_{1,2}^{\psi, p r} \rangle )
& \bar{g}_{2}=g_{2} \sqrt{g}^{11}=g_{2} \sqrt{g}^{22},
\end{align}
where $\langle \cdots \rangle$ indicates we've taken the vacuum expectation value, and $\sqrt{g}^{11}$, etc. are short for $\langle \sqrt{g^{-1}} \rangle_{AB}$. Here, $\bar g_2$ (compared to the SM Lagrangian parameter $g_2$) encapsulates the change in normalization of $W$ in the presence of operators like $H^{\dag}HW^{I\mu\nu} W^I_{\mu\nu}$, while $L^{\psi}_{1,1}, L^{\psi}_{1,2}$ contain the $ffW$ pieces from operators in Eq~\eqref{eq:d6type2}. In addition to being convenient, the geoSMEFT organization is a further step in the geometric picture of effective field theory. First advocated in the context of SMEFT in Ref.~\cite{Alonso:2015fsp, Alonso:2016oah}, the metrics in Eq.~\ref{eq:geosmeft} can be interpreted as interactions on a curved manifold in field space.

Using our baseline flavor assumption, the dimension six four fermion operators that can contribute to monolepton and dilepton production are:
\begin{align}
\label{eq:dim6cont}
& C^{(6)}_{LQ}(L^{\dag}\bar \sigma^\mu L)(Q^{\dag}\bar \sigma_\mu Q)+ C^{3,(6)}_{LQ}(L^{\dag}\bar \sigma^\mu \tau^I L)(Q^{\dag}\bar \sigma_\mu\,\tau^I Q) + C^{(6)}_{Lu}(L^{\dag}\bar \sigma^\mu L)(u^{\dag}_c \bar{\sigma}_\mu u_c) \nn
&+ C^{(6)}_{Ld}(L^{\dag}\bar \sigma^\mu L)( d^{\dag}_c \bar \sigma_\mu d_c ) +  C^{(6)}_{eQ}(e^{\dag}_c \bar \sigma^\mu e_c)(Q^{\dag}\bar \sigma_\mu Q) +  C^{(6)}_{eu}( e^{\dag}_c \bar \sigma_\mu e_c)(  u^{\dag}_c \bar \sigma^\mu u_c )\nn
& ~~~~~~~~~~~~~~~~~~~~~~~~~~ + C^{(6)}_{ed}( e^{\dag}_c \bar\sigma_\mu e_c)(d^{\dag}_c \bar \sigma^\mu d_c),
\end{align}
where we have also dropped all flavor indices as these are trivial with our flavor assumption. Relaxing the flavor symmetry to $U(3)_Q \times U(3)_{u+d} \times U(3)_{L}\times U(3)_{e}$ does not change the counting, while relaxing to $U(3)_{Q+u+d} \times U(3)_{L+e}$ adds terms $C^{(6)}_{LQeu} \epsilon^{ik} L_i Q_k\, e_c\, u_c + h.c.$ and $C^{(6)}_{LQed}\, L^{\dag}_i Q^i e^{\dag}_c d_c$. 

The impact on mono- and dilepton physics from dimension six operators alone has been studied extensively in the literature~\cite{Falkowski:2014tna, Berthier:2015gja, Berthier:2015oma, Falkowski:2015krw, Farina:2016rws,Falkowski:2017pss,Alte:2018xgc,Dawson:2018jlg,Torre:2020aiz,Panico:2021vav,Madigan:2021uho,Dawson:2021ofa}, both at leading order and at next-to-leading order, and the operators here are included in standard SMEFT Monte Carlo packages~\cite{Brivio:2017btx,2021}.

\subsection{Operators at dimension eight}
\label{sec:opsdim8}
At dimension eight, the number of operator classes balloons\footnote{Complete SMEFT bases to dimension eight have been presented in Ref.~\cite{Murphy:2020rsh,Li:2020gnx}. The basis presented here is similar to Ref.~\cite{Li:2020gnx}, as the derivative terms are constructed using the same strategy, however we have massaged the operators into products of currents rather than products of scalars to facilitate combing them with the SM.}. However, many classes can be dismissed right away as they do not fit into any of the bullet points mentioned in Sec.~\ref{sec:intro}. Using the notation of \cite{Grzadkowski:2010es}, where $\psi$ refers to any fermion, $X$ to any field strength, $H$ for Higgs or its conjugate, and $D$ a covariant derivative that can be applied to any field, incompatibility with the bullet points removes operator classes $\psi^4X$, $\psi^2\, H\, X^2$, $H^3\, X^2$, $X^4$, $D\psi^2\, X^2$, $D^2 H^2\,X^2$, $D^2 H^4\,X$, $D\psi^2\, H^2\, X$, $D^2\psi^2\,H\,X$, $D^3\psi^2H^2$, $D^4H^4$, $D^2\psi^2H$ and $D\psi^4H$. For some of these operator types, e,g, $\psi^4X$, the incompatibility is obvious, while for other types it is a bit more subtle. The subtle cases involve operators that, from their field content alone, look like they could contribute to the three-point $\bar f f V$ vertices, such as $D^3\psi^2H^2$. However, as a result of the geoSMEFT organization, the operators that can possibly contribute to two and three-particle vertices is known to all orders, and none of the operators in this category fall into this list. What this means is that, even if an operator  appears to contribute to  $\bar f f V$  (or any other two/three-particle interaction) from its field content, we can always partition the derivatives in such a way that the operators contribute to only four particle and higher interactions. Explicitly, for the $D^3\psi^2H^2$ example, this tells us that at least two of the derivatives can always by manipulated by the equations of motion (EOM) and integration by parts (IBP) to sit on the Higgses, so that all resulting vertices involve two fermions and either two Higgses, two electroweak (EW) gauge bosons, or one Higgs and one EW gauge boson. None of these vertices can interfere with the SM mono/dilepton amplitude (at tree level). The choice of where to put derivatives constitutes a basis, so the counting we are performing here is specific to a those dimension eight operator bases that are `geoSMEFT compatible', meaning that derivatives have been placed to minimize the number of operators contributing at any given order to two and three-particle vertices.

As dimension eight operators enter $\mathcal O(1/\Lambda^4)$ through interference with the SM, we can further neglect all operators whose helicity or flavor structure does not match the SM. Looking at the helicity structure first, tree-level SM amplitudes for dilepton production are products of fermionic currents of the form $(\bar L L)( \bar L L)$,  $(\bar R R)( \bar L L)$ and $(\bar R R)( \bar R R)$, where $L/R$ refer to the fermion helicity. For monolepton production, only left handed fermions participate. Operators that yield $(\bar L R)$ helicity structure -- such as $\psi^2 H^5$ -- the analog of the SM Yukawa term carrying additional Higgses, $\psi^2 H^3 X$  (the dipole terms), and a subset of $\psi^4 H^2$ and $D^2\psi^4$ terms -- cannot interfere and are eliminated.

Taking these two restrictions into consideration, we are left with only five operator structures: $H^4X^2$, $D^2H^6$, $D\psi^2H^4$, $\psi^4 H^2$ and $D^2\psi^4$. 
The first two structures contain the dimension eight contributions to the $h_{IJ}$  and $g_{AB}$ metrics and are independent of any flavor assumptions:
\begin{align}
& C_{H D}^{(8)}(H^{\dagger} H)^{2}\left(D_{\mu} H\right)^{\dagger}\left(D^{\mu} H\right)+C_{H, D 2}^{(8)}(H^{\dagger} H)(H^{\dagger} \tau_I H)\left(D_{\mu} H\right)^{\dagger} \tau^{I}\left(D^{\mu} H\right) \nn
& \quad \quad +C^{(8)}_{HB}\, (H^{\dag}H)^2\, B^{\mu\nu}B_{\mu\nu} + C^{(8)}_{HW}\, (H^{\dag}H)^2\, W^{I,\mu\nu}W_{I,\mu\nu} \nn
& + C^{(8)}_{HW,2}(H^{\dag}\tau^I H)(H^{\dag}\tau^J H)W^{I,\mu\nu}W^J_{\mu\nu} + C^{(8)}_{HWB}\, (H^{\dag}H)(H^{\dag}\tau_I H)\, W^{I,\mu\nu}B_{\mu\nu}.
\end{align}
Similarly, the $D\psi^2H^4$ operators are the dimension eight contribution to $L^{\psi}_{I,A}(\phi)$:
\begin{align}
\label{eq:d8type2}
& C^{1,(8)}_{H\psi} i (H^\dag H)(H^{\dag}\overleftrightarrow D_\mu H)\, (\psi^{\dag}_i \bar{\sigma}^\mu \psi_j),\quad \psi \in \{Q,u_c, d_c, L, e_c\} \nn
& C^{2,(8)}_{H\psi} i (H^\dag \tau^I H)(H^{\dag}\overleftrightarrow D_\mu H)\, (\psi^{\dag}_i \bar{\sigma}^\mu\tau_I \psi_j),\quad \psi \in \{Q, L\}, \nn
& C^{3,(8)}_{H\psi} i (H^\dag H)(H^{\dag}\overleftrightarrow D^I_\mu H)\, (\psi^{\dag}_i \bar{\sigma}^\mu\tau_I \psi_j),\quad \psi \in \{Q, L\}, \nn
& C^{\epsilon,(8)}_{H\psi} i\, \epsilon_{IJK}\, (H^\dag\,\tau^K H)(H^{\dag}\overleftrightarrow D^J_\mu H)\, (\psi^{\dag}_i \bar{\sigma}^\mu\tau^I \psi_j),\quad \psi \in \{Q, L\}, \nn
& \textcolor{blue}{C^{(8)}_{Hud}\, i (H^\dag H)(\tilde H^\dag D_\mu H) (d^{\dag}_{c,i} \bar\sigma^{\mu} u_{c,i}) + h.c.}
\end{align}
Adding these to the metrics and using Eq.~\eqref{eq:geosmeft}, we can can derive the dimension eight contributions to the SM couplings and masses. We have written the fermions in Eq.~\eqref{eq:d8type2} with different flavor indices. However, the combinations with $i \ne j$ cannot interfere with the SM, so only the diagonal entries -- 3 operators per field type  for each operator-- can contribute at $\mathcal O(1/\Lambda^4)$, regardless of the flavor structure of the original operators. Imposing our baseline flavor symmetry sets the coefficients equal for all generations (and forbids the right handed current terms), leaving us with only one operator per field type.

The remaining two structures are four fermion terms. The $\psi^4 H^2$ are the analog of Eq.~\eqref{eq:d6type2}, dressed with an additional power of $H^\dag H$. Given that the Higgs is an electroweak doublet, the extra power of $H^\dag H$ introduces more electroweak structures in some cases.
\begin{align}
& C^{(8)}_{Hed}\, (H^{\dag}H)( e^\dag_c \,\bar \sigma^\mu\, e_c)(d^\dag_c\, \bar\sigma_\mu\, d_c) + C^{(8)}_{Heu}\,  (H^{\dag}H)(e^\dag_c \,\bar\sigma^\mu\, e_c)(u^\dag_c\, \bar{\sigma}_\mu\, u_c) + \nn
& C^{1,(8)}_{HLu}\,  (H^{\dag}H)(L^{\dag}\, \bar{\sigma}^\mu\, L)(u^\dag_c\, \bar{\sigma}_\mu\, u_c)  + C^{3,(8)}_{HLu}\,  (H^{\dag}\tau^I\,H)(L^{\dag}\, \bar{\sigma}^\mu\tau_I\, L)(u^\dag_c\, \bar{\sigma}_\mu\, u_c) + \nn
& C^{1,(8)}_{HLd}\,  (H^{\dag}H)(L^{\dag}\, \bar{\sigma}^\mu\, L)(d^\dag_c\, \bar{\sigma}_\mu\, d_c)  + C^{3,(8)}_{HLd}\,  (H^{\dag}\tau^I\,H)(L^{\dag}\, \bar{\sigma}^\mu\tau_I\, L)(d^\dag_c\, \bar{\sigma}_\mu\, d_c) + \nn
& C^{1,(8)}_{HeQ}\,  (H^{\dag}H)(e^\dag_c\, \bar{\sigma}^\mu\, e_c)(Q^{\dag}\, \bar{\sigma}_\mu\, Q)  + C^{3,(8)}_{HeQ}\,  (H^{\dag}\tau^I\,H)(e^\dag_c\, \bar{\sigma}^\mu\, e_c)(Q^{\dag}_c\, \bar{\sigma}_\mu\,\tau_I Q) + \nn
&  C^{1,(8)}_{HLQ}\,  (H^{\dag}H)(L^{\dag}\, \bar{\sigma}^\mu\, L)(Q^{\dag}\, \bar{\sigma}_\mu\, Q) + C^{2,(8)}_{HLQ}\,  (H^{\dag}H)(L^{\dag}\, \bar{\sigma}^\mu\,\tau^I\, L)(Q^{\dag}\, \bar{\sigma}_\mu\,\tau_I Q) +\nn
& C^{3,(8)}_{HLQ}\,  (H^{\dag}\tau^I\, H)(L^{\dag}\, \bar{\sigma}^\mu\,\tau_I L)(Q^{\dag}\, \bar{\sigma}_\mu\, Q) + C^{4,(8)}_{HLQ}\,  (H^{\dag}\tau^I\,H)(L^{\dag}\, \bar{\sigma}^\mu\, L)(Q^{\dag}\, \bar{\sigma}_\mu\,\tau_I Q) + \nn
&  C^{5,(8)}_{HLQ}\, \epsilon_{IJK}  (H^{\dag}\tau^I\,H)(L^{\dag}\, \bar{\sigma}^\mu\,\tau^J\, L)(Q^{\dag}\, \bar{\sigma}_\mu\,\tau^K\, Q) +  \textcolor{blue}{C^{(8)}_{HLud} i H^2 (L^{\dag}\bar\sigma^\mu L)(d^\dag_c \bar\sigma_\mu u_c) + h.c.} 
\label{eq:contactH1}
\end{align}
Dressing the fields with flavor indices, interference with the SM projects out only the diagonal flavor structures, e.g. operators of the form $(\psi^{\dag}_i\psi_i)(\chi^\dag_j \chi_j)$ where $\psi, \chi$ are SM fermions. $(U(3))^5$ reduces the number from 9, 3 for $i$ times 3 for $j$ for each $\psi$ and $\chi$, to 1. Loosening the flavor symmetry to $U(3)_Q\times U(3)_{d+u} \times U(3)_{L}\times U(3)_{e}$ allows one extra structure (in blue). Further relaxing the symmetry to $U(3)_{Q+d+u} \times U(3)_{L+e}$ we can write operators containing $d_c\, Q\,e_c\, L\, (H^\dag)^2, d_c\, Q\, e^{\dag}_c\, L^\dag (H^\dag H), u_c\, Q\, e_c\, L\, (H^\dag H)$ and $u_c\, Q\, e_c\, L\, H^2$ (all requiring h.c.), however none of these have the right helicity structure to interfere with the SM (in the limit of massless fermions) so they cannot enter at $\mathcal O(1/\Lambda^4)$.

Of the 14 operator structures written in Eq.~\eqref{eq:contactH1}, only two -- $C^{2,(8)}_{HLQ}$ and $C^{5,(8)}_{HLQ}$, which involve both left handed lepton and quark EW triplets -- lead to products of charged currents and will enter into monolepton production\footnote{ The operator $C^{(8)}_{HLud} i H^2 (L^{\dag}\bar\sigma^\mu L)(d^\dag_c \bar\sigma_\mu u_c) + h.c.$ also generates products of charged currents. However, even if it is permitted by the flavor symmetry, there is no SM right handed quark current for this operator to interfere with so it does not appear at $\mathcal O(1/\Lambda^4)$.}. The $C^{5,(8)}_{HLQ}$ structure actually exclusively generates products of charged currents and therefore cannot play a role in dilepton production, while the remaining 13 will enter.

The second type of dimension eight, four fermion terms have the form $D^2\psi^4$. At first glance, the counting seems trickier here as we have the choice to put the two derivatives on any of the four fields (so six possible arrangements), but many of these choices are related through EOM or IBP. This is best seen in momentum space, where the derivative $D\psi_i \to p_i$\footnote{The derivatives here are covariant derivatives, so $D\psi$ contains pieces proportional to $A\, \psi$, where $A$ is a gauge field. The terms with gauge fields do not contribute to dilepton production, however, so for our purposes we only care about the ordinary derivative piece of $D$.}, and redundancies from EOM and IBP are more straight forward. Specifically, by the EOM we can remove terms where both derivatives hit the same field, all $p^2_i$. So, for two derivatives and four fields -- which we label $\psi_{1 .. 4}$ -- we are left with the combinations $s_{12}$, $s_{13}$, $s_{14}$, $s_{23}$, $s_{24}$, $s_{34}$, where $s_{ij} = 2p_i\cdot p_j$. Momentum conservation (IBP in momentum space) allows us to remove one of the $p_i$ in favor of the others. Choosing $p_4$, this reduces the set of invariants to   $s_{12}$, $s_{13}$ and $s_{23}$ -- which correspond to the usual Mandelstam variables $s, t, u$. One of these three can be removed using $s + t + u = 0$, leaving only the two invariants; for operators admitting multiple electroweak structures, we will have two $D^2$ operators for each structure. Translated back to operators, and choosing the two momentum combinations to be Mandelstam $s$ and $t$ -- yet another basis choice -- we have\footnote{The coefficients of all the $D^2$ are real, hence their placement outside the parenthesis for the $C^{t,(8)}$ operators.}:
\begin{align}
&  C^{s,(8)}_{ed}\, (e^\dag_c\, \bar{\sigma}^{\mu}\, e_c)D^2 (d^\dag_c\, \bar{\sigma}_{\mu}\, d_c) + C^{t,(8)}_{ed}\, \Big((D_\nu e^\dag_c\, \bar{\sigma}^{\mu}\,  e_c)(D^{\nu}d^\dag_c\,\bar {\sigma}_{\mu}\, d_c) + h.c.\Big)+ \nn
&  C^{s,(8)}_{eu}\, (e^\dag_c\, \bar{\sigma}^{\mu}\, e_c)D^2(u^\dag_c\,\bar {\sigma}_{\mu}\, u_c) + C^{t,(8)}_{eu}\, \Big((D_\nu e^\dag_c\, \bar{\sigma}^{\mu}\,  e_c)(D^{\nu} u^\dag_c\, \bar{\sigma}_{\mu}\, u_c) + h.c. \Big) + \nn
&  C^{s,(8)}_{eQ}\, (e^\dag_c\, \bar{\sigma}^{\mu}\, e_c)D^2(Q^{\dag}\, \bar{\sigma}_{\mu}\, Q) + C^{t,(8)}_{eQ}\, \Big((D_\nu e^\dag_c\, \bar{\sigma}^{\mu}\,  e_c)(D^{\nu} Q^{\dag}\, \bar{\sigma}_{\mu}\, Q) + h.c. \Big) +\nn
&  C^{s,(8)}_{Ld}\, (L^{\dag}\, \bar{\sigma}^{\mu}\, L)D^2(d^\dag_c\, \bar{\sigma}_{\mu}\, d_c) + C^{t,(8)}_{Ld}\, \Big((D_\nu L^{\dag}\, \bar{\sigma}^{\mu}\,  L)(D^{\nu}d^\dag_c\, \bar{\sigma}_{\mu}\, d_c) +  h.c. \Big) + \nn
&  C^{s,(8)}_{Lu}\, (L^{\dag}\, \bar{\sigma}^{\mu}\, L)D^2(u^\dag_c\, \bar{\sigma}_{\mu}\, u_c) + C^{t,(8)}_{Lu}\, \Big((D_\nu L^{\dag}\, \bar{\sigma}^{\mu}\,  L)(D^{\nu}u^\dag_c\, \bar{\sigma}_{\mu}\, u_c) + h.c. \Big) + \nn
&  C^{s,1,(8)}_{LQ}\, (L^{\dag}\, \bar{\sigma}^{\mu}\, L)D^2(Q^{\dag}\, \bar{\sigma}_{\mu}\, Q) + C^{t,1,(8)}_{LQ}\, \Big((D_\nu L^{\dag}\, \bar{\sigma}^{\mu}\,  L)(D^{\nu}Q^{\dag}\, \bar{\sigma}_{\mu}\, Q) +h.c. \Big) + \nn
&  C^{s,3,(8)}_{LQ}\, (L^{\dag}\, \bar{\sigma}^{\mu}\,\tau^I L)D^2(Q^{\dag}\, \bar{\sigma}_{\mu}\,\tau_I Q) + C^{t,3,(8)}_{LQ}\, \Big((D_\nu L^{\dag}\, \bar{\sigma}^{\mu}\,\tau^I  L)(D^{\nu}Q^{\dag}\, \bar{\sigma}_{\mu}\,\tau_I Q) + h.c. \Big).
\label{eq:contactH3}
\end{align}
The momentum space perspective for counting operators has been advocated as part of the `on-shell' approach, see Ref.~\cite{Shadmi:2018xan,Ma:2019gtx,Henning:2019enq, Henning:2019mcv}. Dressing the derivative operators with flavor indices, the story is the same as with four fermion operators dressed with Higgses.  The only operators that survive interference with the SM are flavor diagonal, so 9 operators per structure appearing in Eq.~\eqref{eq:contactH3}. This is reduced to one per structure under our baseline flavor symmetry. Reducing the flavor symmetry to $U(3)_{Q+u+d} \times U(3)_{L+e}$ has no effect at $\mathcal O(1/\Lambda^4)$ as the additional four fermion terms that are allowed do not interfere with the SM\footnote{No additional operators are generated if the flavor symmetry is only reduced to  $U(3)_{Q}\times U(3)_{u+d} \times U(3)_{L+e}$. }.

Only the two structures involving the contraction of lepton and quark EW triplets, $C^{s,3,(8)}_{LQ}$ and $C^{t,3,(8)}_{LQ}$ will have the correct field content to interfere with SM monolepton production. All operator structures can participate in dilepton processes. 

While we are interested in analytic results for our purposes and will give analytic expressions in later sections, incorporating dimension eight operators into Monte Carlo programs is not as onerous as it may seem. The dimension eight contributions to the $h_{IJ}, g_{AB}$, $L^{\psi}_{I,A}$ metrics and the four fermion terms with additional Higgses (Eq.~\eqref{eq:contactH1}) do not introduce kinematic structures beyond what's present at dimension six and can thereby be accounted for (numerically, at least) by rescaling Monte Carlo dimension six results. Only the dimension eight, four fermion terms with extra derivatives require modifications to code, e.g. SMEFTsim~\cite{Trott:2021vqa}.

\subsection{Helicity amplitude expansion for monolepton production}
\label{sec:helampexpand}

Now that we have enumerated the important operators at dimension six and eight, we can work out the helicity amplitudes. As a first step, we look at the pieces coming from $W$ exchange (for monolepton) or photon/$Z$ exchange (for dilepton). These contributions are functions of $ffV$ couplings and the gauge boson masses and widths, all of which have a SM value but receive corrections at each order in $1/\Lambda$ from operators lying in the $h_{IJ}, g_{AB}$ and $L^{\psi}_{I,A}$ metrics. The corrections are well defined within a given operator basis. However, we have a choice in how to connect these quantities to experiment, specifically through what observables are used to define the electroweak theory inputs $g_1, g_2$ and $v$. In the $\hat{\alpha}_{ew}$ scheme, $G_F$ fixes $v$, while the electromagnetic coupling and $m_Z$ set $g_1, g_2$, while in the $\hat{m}_W$ scheme, $G_F$ fixes $v$ and $m_W, m_Z$ fix the gauge couplings.

We will present our results here in the $\hat m_W$ scheme, deferring the $\hat\alpha_{ew}$ scheme to Appendix~\ref{sec:alphascheme}. Choosing the $\hat m_W$ scheme means that the electromagnetic coupling is a derived quantity and therefore is a function of SMEFT inputs. The reward for this inconvenience is that $m_W$ is an input, so expressions for monolepton production will be simpler. Schematically carrying out the SMEFT expansion for the electromagnetic coupling, $W/Z$ couplings and widths, we have  
\begin{align}
\label{eq:expand1}
g^{eff}_{W, f_{L}}  = g^{0}_{W, f_{L}} + x\,  \delta g^{(1)}_{W, f_{L}} + x^2\,  \delta g^{(2)}_{W, f_{L}} + \cdots, \quad & e_{em} = e^{0} + x\, \delta e^{(1)} + x^2\, \delta e^{(2)} + \cdots \nn
g^{eff}_{W, q_{R}}  =  \textcolor{blue}{x\,  \delta g^{(1)}_{W, q_{R}} + x^2\,  \delta g^{(2)}_{W, q_{R}} + \cdots}, \quad & g^{eff}_{Z, f_{L/R}}  = g^{0}_{Z, f_{L/R}} + x\,  \delta g^{(1)}_{Z, f_{L/R}} + x^2\,  \delta g^{(2)}_{Z, f_{L/R}} + \cdots  \nn
g^{eff}_{W, \ell_{R}} = 0  \quad\quad\quad \quad\quad\quad& \quad\quad\quad \nn
\Gamma_W = \Gamma_{W,0} + x\,\delta \Gamma^{(1)}_{W} + x^2 \,\delta \Gamma^{(2)}_{W} + \cdots, \quad & \Gamma_Z = \Gamma_{Z,0} + x\,\delta \Gamma^{(1)}_{Z} + x^2 \,\delta \Gamma^{(2)}_{Z} + \cdots\,
\end{align}
where $x = \frac{v^2_T}{\Lambda^2}$. As before, blue text indicates corrections that are absent under our baseline flavor symmetry but appear if we loosen the flavor symmetry to $U(3)_Q\times U(3)_{u+d}\times U(3)_{L}\times U(3)_{e}$. As there is no right handed neutrino in the SMEFT paradigm, the $W$ coupling to right handed leptons is zero to all orders in the $x$ expansion, while the $W$ coupling to right handed quarks is nonzero starting at $\mathcal O(x)$, provided we have assumed a flavor symmetry loose enough to permit it.

Here, $ \delta e^{(1,2)}$, $\delta g^{(1,2)}_{W, f_{L/R}}, \delta g^{(1,2)}_{Z, f_{L/R}}$ etc. are functions of Wilson coefficients. This dependency has been worked out for the $Z$ couplings and width in Ref.~\cite{Hays:2020scx, Corbett:2021eux} using the neutral current version of Eq.~\eqref{eq:geosmeft} expanded to $\mathcal O(x^2)$ (for both electroweak input schemes). While the corrections to the $W$ couplings $\delta g^{(i)}_{W, f_{\lambda_f}}$ can be extracted from those references, for convenience we have listed them along with the corrections to $\Gamma_W$:
\begin{align}
\label{eq:coeff1}
& e^0 = 0.308 \\
& \delta e^{(1)} = -0.576\, C^{(6)}_{HWB} - 0.269\, C^{(8)}_{HD} - 0.218\, \delta G^{(6)}_F \nn
& \delta e^{(2)} = -0.288\, C^{(8)}_{HWB} - 0.0385\, C^{(8)}_{HD} - 0.230\, C^{(8)}_{HD,2} - 0.576\, C^{(6)}_{HB}\,C^{(6)}_{HWB} - 0.144\, C^{(6)}_{HD}\, C^{(6)}_{HWB} \nn
& ~~~~~~~~~ - 0.576\, C^{(6)}_{HW}\, C^{(6)}_{HWB} + 0.406\, C^{(6)}_{HWB}\,\delta G^{(6)}_F - 0.117\, (C^{(6)}_{HD})^2 + 0.190\, C^{(6)}_{HD}\, \delta G^(6)_F\nn
& ~~~~~~~~~ + 0.231\, (\delta G^{(6)}_F)^2 - 0.218\, \delta G^{(8)}_F \nonumber 
\end{align}
\begin{align}
\label{eq:coeff2}
& g^0_{W, \ell_L} = -0.46 \\
& \delta g^{(1)}_{W, \ell_L} =  -0.46\, C^{3,(6)}_{HL} + 0.33\, \delta G^{(6)}_F \nn
& \delta g^{(2)}_{W, \ell_L} =  0.33\,  C^{3,(6)}_{HL}\, \delta G^{(6)}_F - 0.35 (\delta G^{(6)}_F)^2 - 0.23\, C^{3,(8)}_{HL} + 0.058\, C^{(8)}_{HD,2} + 0.23\, i\, C^{(8)}_{\epsilon H} + 0.33\,\delta G^{(8)}_F \nn
& g^0_{W, \ell_R}  = \delta g^{(1)}_{W, \ell_R}  = \delta g^{(2)}_{W, \ell_R} = 0  \nonumber
\end{align}
\begin{align}
\label{eq:coeff3}
& g^0_{W, f_L} = -0.47 \\
& \delta g^{(1)}_{W, f_L} =  -0.46\, C^{3,(6)}_{HQ} + 0.33\, \delta G^{(6)}_F \nn
& \delta g^{(2)}_{W, f_L} =  0.33\,  C^{3,(6)}_{HQ}\, \delta G^{(6)}_F - 0.35 (\delta G^{(6)}_F)^2 - 0.23\, C^{3,(8)}_{HQ} + 0.058\, C^{(8)}_{HD,2} + 0.23\, i\, C^{(8)}_{\epsilon H} + 0.33\,\delta G^{(8)}_F \nn
& g^0_{W, f_R}  = 0 \nn
& \delta g^{(1)}_{W, f_R}  =  \textcolor{blue}{- 0.23 i C^{(6)}_{Hud}}\nn
& \delta g^{(2)}_{W, f_R} = \textcolor{blue}{0.16 i C^{(6)}_{Hud}\, \delta G^{(6)}_F - 0.12i\, C^{(8)}_{Hud}}  \nonumber
\end{align}
Finally, for the $W$ width: 
\begin{align}
\label{eq:coeff4}
&\Gamma^{0}_W\, (\textrm{GeV}) = 2.05\, \\
& \delta \Gamma^{(1)}_W\, (\textrm{GeV}) = 1.36\, C^{3,(6)}_{HL} + 2.73\, C^{3,(6)}_{HQ} - 2.89\,  \delta G^{(6)}_F\nn
& \delta \Gamma^{(2)}_W\ (\textrm{GeV}) = 0.682\, (C^{3,(6)}_{HL} )^2 -1.93\, C^{3,(6)}_{HL} \delta G^{(6)}_F + 1.36\, (C^{3,(6)}_{HQ})^2 - 3.86\, C^{3,(6)}_{HQ}\, \delta G^{(6)}_F \nn
& ~~~~~~~~~~~~~ \textcolor{blue}{+ 0.341\, (C^{(6)}_{Hud})^2} + 4.09\,( \delta G^{(6)}_F)^2 + 0.682\, C^{3,(8)}_{HL} +1.36\,  C^{3,(8)}_{HQ} -  \nn
& ~~~~~~~~~~~~-0.511\, C^{(8)}_{HD} + 0.511\, C^{(8)}_{HD,2}  - 2.89\, \delta G^{(8)}_F \nonumber
\end{align}
 Note that, had we done things in the $\hat \alpha_{ew}$ scheme, $e_{em}$ would be an input and would have no $x$ expansion, while we would have to add $m^{eff}_W = m^0_W + x\, \delta m^{(1)}_W + x^2\,\delta m^{(2)}_W + \cdots$ to the list. 

Moving to the contact/four fermion terms, we have seen they come in two varieties, i.) accompanied by powers of $(H^{\dag}H)$, which become powers of $\bar v^2_T$, and ii.) accompanied by derivatives, which become factors of Mandelstam $s$ or $t$. Allowing both of these forms, we can express $a^{eff}_{\lambda_f,\lambda_\ell}$ as
\begin{align}
a^{eff}_{\lambda_f,\lambda_\ell} = \frac{x}{\bar v^2_T}\, \delta a^{(1)}_{\lambda_f,\lambda_\ell} + \frac{x^2}{\bar v^2_T}\,\Big( \delta a^{(2)}_{\lambda_f,\lambda_\ell}  - \frac{\hat s}{\bar v^2_T} \delta a^{(2,s)}_{\lambda_f,\lambda_\ell} - \frac{\hat t}{\bar v^2_T} \delta a^{(2,t)}_{\lambda_f,\lambda_\ell} \Big).
\label{eq:expand2}
\end{align}
It is straight forward to map the contact operators in Eq.~\eqref{eq:dim6cont}, \eqref{eq:contactH1} and \eqref{eq:contactH3} into this format. We will group the terms according to whether they contribute to dilepton production (using a superscript $NC$ for neutral current) or monolepton production (using superscript $CC$ for charged current). At dimension six:
\begin{align}
 \delta a^{NC,(1)}_{LL} =C^{(6)}_{LQ} \mp C^{3,(6)}_{LQ},\quad &\delta a^{NC, (1)}_{LR} = C^{(6)}_{eQ},\quad \delta a^{NC,(1)}_{RL} = \left\{\begin{array}{c} C^{(6)}_{Lu} \\ C^{(6)}_{Ld} \end{array} \right.,\quad\delta a^{NC,(1)}_{RR} = \left\{\begin{array}{c} C^{(6)}_{eu} \\ C^{(6)}_{ed} \end{array} \right.\nn
& \delta a^{CC,(1)}_{LL} = 2\,C^{3,(6)}_{LQ},\quad  \delta a^{CC,(1)}_{RL} = 0, 
\end{align}
where the upper/lower sign of $\mp$ refers to up/down quarks. At dimension eight:
\begin{align}
& \delta a^{NC,(2)}_{LL} = \frac{C^{1,(8)}_{HLQ}}{2} \mp \frac{C^{2,(8)}_{HLQ}}{2} + \frac{C^{3,(8)}_{HLQ}}{2} \mp \frac{C^{4,(8)}_{HLQ}}{2} \quad \nonumber \\
& \delta a^{NC,(2,s)}_{LL} = \frac{C^{s,1,(8)}_{LQ}}{2} \mp \frac{C^{s,3,(8)}_{LQ}}{2},\quad \delta a^{NC,(2,t)}_{LL} = \frac{C^{t,1,(8)}_{LQ}}{2} \mp \frac{C^{t,3,(8)}_{LQ}}{2}\nn
& \delta a^{NC,(2)}_{LR} = \frac{C^{1,(8)}_{HeQ}}{2} \mp \frac{C^{3,(8)}_{HeQ}}{2}, \quad \delta a^{NC,(2,s)}_{LR} = C^{s,(8)}_{eQ},\quad \delta a^{NC,(2,t)}_{LR} = C^{t,(8)}_{eQ} \nn
& \delta a^{NC,(2)}_{RL} = \left\{ \begin{array}{c} \frac{C^{1,(8)}_{HLu}}{2} + \frac{C^{3,(8)}_{HLu}}{2} \\ \frac{C^{1,(8)}_{Ld}}{2} + \frac{C^{3,(8)}_{HLd}}{2} \end{array}\right.
\quad, \delta a^{NC,(2,s)}_{RL} = \left\{ \begin{array}{c} C^{s,(8)}_{Lu} \\  C^{s,(8)}_{Ld} \end{array} \right.,\quad \delta a^{NC,(2,t)}_{RL} =  \left\{ \begin{array}{c} C^{t,(8)}_{Lu} \\  C^{t,(8)}_{Ld} \end{array} \right.\nn
&\delta a^{NC,(2)}_{RR} = \left\{ \begin{array}{c} \frac{C^{(8)}_{Heu}}{2} \\ \frac{C^{(8)}_{Hed}}{2} \end{array} \right., \quad \delta a^{NC,(2,s)}_{RR} =  \left\{ \begin{array}{c} C^{s,(8)}_{eu} \\ C^{s,(8)}_{ed} \end{array} \right. ,\quad \delta a^{NC,(2,t)}_{RR} =  \left\{ \begin{array}{c} C^{t,(8)}_{eu} \\ C^{t,(8)}_{ed} \end{array} \right.
\end{align}

\begin{align}
& \delta a^{CC,(2)}_{LL} =C^{2,(8)}_{HLQ} \mp i C^{5,(8)}_{HLQ}\,,\quad \delta a^{CC,(2,s)}_{LL} =2\,C^{s,3,(8)}_{LQ} \,,\quad \delta a^{CC,(2,t)}_{LL} =2\,C^{t,3,(8)}_{LQ}, 
\end{align}
where the upper (lower) sign in $\delta a^{CC,(2)}_{LL} $ refers to $W^+\, (W^-)$ production. 

Combining the tree-exchange and contact pieces together, we can write the full helicity amplitudes as an expansion in $x$
\begin{align}
\mathcal A^{tot}_{\lambda_f,\lambda_\ell} = \mathcal A^{(0)}_{\lambda_f,\lambda_\ell}  +  x\, \mathcal A^{(1)}_{\lambda_f,\lambda_\ell}  + x^2  \mathcal A^{(2)}_{\lambda_f,\lambda_\ell}  + \cdots
\end{align}
where the real and imaginary pieces of $ \mathcal A^{(i)}_{\lambda_f,\lambda_\ell} $ are functions of the parameters in Eq.~\eqref{eq:expand1},~\eqref{eq:expand2}. As both left and right-handed fermions couple to the $Z/\gamma$ the helicity amplitudes for $pp \to \ell^+\ell^-$ all have a similar structure so it makes sense to present them for general $\lambda_i, \lambda_f$. These expressions are presented in Appendix~\ref{sec:helampNC}. However, for monolepton production, left and right handed fermions couple very differently. The root of the difference in couplings is the fact that there is no right handed neutrino in SMEFT. Therefore, there can be no $W$ coupling to right handed leptons, nor can there be any contact terms for the $(LR)$ or $(RR)$ helicity structures. With these restrictions, the real and imaginary parts of the amplitude are given below to each order in $x$ (in the $\hat m_W$ scheme).  

At $\mathcal O(x^0)$, the only nonzero piece comes from $(\lambda_i\lambda_f) = (LL)$ helicity
\begin{align}
& \text{Re}(\mathcal A^{W,(0)}_{L,L}) = \Big(\frac{(\hat s-m^2_W)\, g^0_{W, f_L}\, g^0_{W, \ell_L}}{P_0(\hat s, m_W, \Gamma^0_W)} \Big), \quad \text{Im}(\mathcal A^{W,(0)}_{L,L}) = -\frac{\Gamma^0_W\, m_W\, g^0_{W, f_{L}}\, g^0_{W, \ell_L} }{P_0(\hat s, m_W, \Gamma^0_W)} 
\end{align}
with $P_0(\hat s, m_W, \Gamma^0_W) = (\hat s - m^2_W)^2 + (\Gamma^0_W)^2m^2_W$. \\

At $\mathcal O(x)$, $\mathcal A^{(1)}_{LL},  \mathcal A^{(1)}_{RL}$ are nonzero, while $\mathcal A^{(1)}_{LR}, \mathcal A^{(1)}_{RR}$ remain zero:
\begin{align}
\label{eq:helampW}
&\text{Re}(\mathcal A^{W,(1)}_{L,L}) = \Big( - \frac{(m^2_W - \hat s)(\delta g^{(1)}_{W, f_{L}} g^0_{W, \ell_{L}} + \delta g^{(1)}_{W, \ell_{L}} g^0_{W, f_{L}}) }{P_0(\hat s, m_W, \Gamma^0_W)}  + \nn
&~~~~~~~~~~~~~~~~~~~~~~~~~~~~~~~~~~~~~~~~~~~~~~~~~~~~~~~~ \frac{2\,\delta \Gamma^{(1)}_W\, \Gamma^0_W\, m^2_W g^0_{W,f_{L}}\,g^0_{W, \ell_{L}} (m^2_W - \hat s)}{P^2_0(\hat s, m_W, \Gamma^0_W)} + \frac{\delta a^{W,(1)}_{L,L}}{\bar v^2_T} \Big) \nn
&\text{Im}(\mathcal A^{W,(1)}_{L,L}) = \Big(- \frac{\Gamma^0_W\, m_W\,(\delta g^{(1)}_{W, f_{L}} g^0_{W, \ell_{L}} + \delta g^{(1)}_{W, \ell_{L}} g^0_{W, f_{L}})  }{P_0(\hat s, m_W, \Gamma^0_W)} -\frac{\delta \Gamma^{(1)}_W\, m_W\, g^0_{W, f_{L}} g^0_{W, \ell_{L}} ((m^2_W - \hat s)^2 - m^2_W\, (\Gamma^0_W)^2)}{P^2_0(\hat s, m_W, \Gamma^0_W )} \Big) \nn
&\text{Re}(\mathcal A^{W,(1)}_{R,L}) = \frac{ g^0_{W, \ell_L} (\text{Re}(\delta g^{(1)}_{W, f_R}) (\hat s - m^2_W) + \text{Im}(\delta g^{(1)}_{W, f_R})\, m_W\, \Gamma^0_W)}{P_0(\hat s, m^2_W, \Gamma^0_W)} + \frac{\delta a^{CC,(1)}_{R,L}}{\bar v^2_T}\nn
&\text{Im}(\mathcal A^{W,(1)}_{R,L}) = -\frac{g^0_{W, \ell_L}(\Gamma^0_W\, m_W\, \text{Re}(\delta g^{(1)}_{W, f_R}) + (m^2_W - s)\text{Im}(\delta g^{(1)}_{W, f_R})  }{P_0(\hat s, m^2_W, \Gamma^0_W)}.
\end{align}

At $\mathcal O(x^2)$, all helicity sub-amplitudes are nonzero. However, for an $\mathcal O(x^2)$ amplitude piece to make a contribution to the cross section at $\mathcal O(x^2)$, it must interfere with the SM ($\mathcal O(x^0))$ piece. As the SM piece has $(LL)$ helicity, the only $\mathcal O(x^2)$  term we care about is  $\mathcal A^{(2)}_{LL}$. One additional complication here is that the $\delta g^{(2)}_{W,f_L}, \delta g^{(2)}_{W,\ell_L}$ and $\delta a^{(2)}_{LL}$ couplings are complex:
\begin{align}
\label{eq:helampW2}
& \text{Re}(\mathcal A^{W,(2)}_{L,L}) =\Big( - \frac{(m^2_W - \hat s)(\delta g^{(1)}_{W, f_{L}} \delta g^{(1)}_{W, \ell_{L}} + \text{Re}(\delta g^{(2)}_{W, f_{L}}) g^0_{W, \ell_{L}} + \text{Re}(\delta g^{(2)}_{W, \ell_{L}}) g^0_{W, f_{L}}  )}{P_0(\hat s, m_W, \Gamma^0_W)} + \nn
&~~~~~~~~~~~~~~~~~~~~~~~~ + \frac{\Gamma^0_W\, m_W (\text{Im}(\delta g^{(2)}_{W, f_{L}}) g^0_{W, \ell_{L}} + \text{Im}(\delta g^{(2)}_{W, \ell_{L}}) g^0_{W, f_{L}})}{P_0(\hat s, m_W, \Gamma^0_W)} \nn
& ~~~~~~~~~~~~~~~~~~~~~~~~~ + \frac{2\delta \Gamma^{(1)}_W\, \Gamma^0_W\, m^2_W\, (m^2_W - \hat s)(\delta g^{(1)}_{W, f_{L}} g^0_{W, \ell_{L}} + \delta g^{(1)}_{W, \ell_{L}} g^0_{W, f_{L}})}{P^2_0(\hat s, m_W, \Gamma^0_W)}  \\
& ~~~~~~~~~~~~~~~~~~~~~~~~~~+ \frac{2\,\delta \Gamma^{(2)}_W\, \Gamma^0_W\, m^2_W\, (m^2_W - \hat s) g^0_{W, f_{L}}g^0_{W, \ell_{L}}}{P^2_0(\hat s, m_W, \Gamma^0_W)}  \nn
& ~~~~~~~~~~~~~~~~~~~~~~+ \frac{(\delta\Gamma^{(1)}_W)^2\, g^0_{W, f_{L}}g^0_{W, \ell_{L}}\, m^2_W\, (m^2_W - \hat s)(m^4_W + \hat s^2 - m^2_W(3(\Gamma^0_W)^2 + 2\hat s) )}{P^3_0(\hat s, m_W, \Gamma^0_W)} + \nn
& ~~~~~~~~~~~~~~~~~~~~~~~~\frac 1 {\bar v^2_T}\Big( \text{Re}(\delta a^{CC,(2)}_{L,L}) - \frac{\hat s }{\bar v^2_T}\delta a^{CC,(2,s)}_{L,L} - \frac{\hat t }{\bar v^2_T}\delta a^{CC,(2,t)}_{L,L}\Big) \nn
& \nn
&\text{Im}(\mathcal A^{W,(2)}_{L,L}) =  \Big( -\frac{\delta \Gamma^{(2)}_W\, m_W\, g^0_{W, f_{L}} g^0_{W, \ell_{L}} ((m^2_W - \hat s)^2 - m^2_W\, (\Gamma^0_W)^2)}{P^2_0(\hat s, m_W, \Gamma^0_W )} \nn
& ~~~~~~~~~~~~~~~~~~~~~~~~~~~~~~~~ -\frac{\Gamma^0_W\,m_W\,(\delta g^{(1)}_{W, f_{L}} \delta g^{(1)}_{W, \ell_{L}} + \text{Re}(\delta g^{(2)}_{W, f_{L}}) g^0_{W, \ell_{L}} + \text{Re}(\delta g^{(2)}_{W, \ell_{L}}) g^0_{W, f_{L}} )}{P_0(\hat s, m_W, \Gamma^0_W)} \nn
&~~~~~~~~~~~~~~~~~~~~~~~~~~~~~~ -\frac{(m^2_W -s)(\text{Im}(\delta g^{(2)}_{W, f_{L}}) g^0_{W, \ell_{L}} + \text{Im}(\delta g^{(2)}_{W, \ell_{L}}) g^0_{W, f_{L}})}{P_0(\hat s, m_W, \Gamma^0_W)} \nn
& ~~~~~~~~~~~~~~~~~~~~~~~~~~~~~ -\frac{\delta\Gamma^{(1)}_W\, m_W\, (\delta g^{(1)}_{W, f_{L}} g^0_{W, \ell_{L}} + \delta g^{(1)}_{W, \ell_{L}} g^0_{W, f_{L}}) (m^4_W + \hat s^2 - m^2_W((\Gamma^0_W)^2 + 2\hat s) )}{P^2_0(\hat s, m_W, \Gamma^0_W)}  \nn
&~~~~~~~~~~~~~~~~~~~~~~~~~~~~~ +\frac{(\delta \Gamma^{(1)}_W)^2\,\Gamma^0_W\, m^3_W(3(m^2_W - \hat s)^2 - (\Gamma^0_W)^2\, m^2_W) g^0_{W, f_{L}} g^0_{W, \ell_{L}} }{P^3_0(\hat s, m_W, \Gamma^0_W)} \Big) \nn
& ~~~~~~~~~~~~~~~~~~~~~~~~~~~~~ +\frac 1 {\bar v^2_T} \text{Im}(\delta a^{CC,(2)}_{L,L}).
\end{align}

In terms of these helicity amplitudes, the full amplitude squared is:
\begin{align}
& \sum_{\lambda_f, \lambda_\ell}|\mathcal A^W_{\lambda_f,\lambda_\ell}|^2 = (\text{Re}(\mathcal A^{W,(0)}_{L,L})^2 + \text{Im}(\mathcal A^{W,(0)}_{L,L})^2 ) + 2x\, (\text{Re}(\mathcal A^{W,(0)}_{L,L})\text{Re}(\mathcal A^{W,(1)}_{L,L}) + \text{Im}(\mathcal A^{W,(0)}_{L,L})\text{Im}(\mathcal A^{W,(1)}_{L,L}) \nn
&  ~~~~~ +  x^2\, \Big(2\, \text{Re}(\mathcal A^{W,(0)}_{L,L})\text{Re}(\mathcal A^{W,(2)}_{L,L}) + 2\, \text{Im}(\mathcal A^{W,(0)}_{L,L})\text{Im}(\mathcal A^{W,(2)}_{L,L}) + \text{Re}(\mathcal A^{W,(1)}_{L,L})^2 + \text{Im}(\mathcal A^{W,(1)}_{L,L})^2 \Big) \nn
& ~~~~~~ + x^2\, \textcolor{blue}{\Big( \text{Re}(\mathcal A^{W,(1)}_{R,L})^2 + \text{Im}(\mathcal A^{W,(1)}_{R,L})^2 \Big)}.
\label{eq:ampsquared}
\end{align}

\section{Parton level result}
\label{sec:Wpparton}

Using the amplitude expressions from Sec.~\ref{sec:helampexpand} and plugging in the coupling factors, $W$ width and contact operators, we can calculate the partonic cross section for monolepton production,
\begin{align}
\hat \sigma (\bar q_i q_j \to \ell^\pm \nu) = \hat \sigma^{(0)}(\hat s) + x\, \hat \sigma^{(1)}(\hat s) + x^2\, \hat \sigma^{(2)}(\hat s) + \cdots
\end{align}
The full expressions for $\hat \sigma^{(0,1,2)}(\hat s)$ are long and not particularly illuminating. To hone in on the subset of dimension six and eight effects that grow with energy the fastest and therefore drive the difference between SMEFT and the SM in high energy experiments, we can take the large $\hat s $ limit:
\begin{align}
\label{eq:leadings}
\hat\sigma_{\hat s \to\infty}(\bar q q' \to \ell^{\pm}\nu) = \frac{\hat s\, x^2}{36\, \pi\, \hat v^4}\Big( (C^{3,(6)}_{LQ})^2 - \frac{\hat e^2}{8\,\hat s^2_\theta}(4C^{s,3,(8)}_{LQ} -  C^{t,3,(8)}_{LQ}) \Big) + \mathcal O(\hat s^0),
\end{align}
where $\hat e, \hat v, \hat s_\theta$ are functions of the measured experimental inputs $\hat m_W, \hat m_Z, \hat G_F$ (in the $\hat m_W$ scheme).\footnote{Explicitly, $\hat v = 1/(2^{1/4}\hat G_F), \hat \theta = \text{sin}^{-1}(\sqrt{1- \hat m^2_W/\hat m^2_Z})\,, \hat e = 2\times 2^{1/4}\, \hat m_W \sqrt{\hat G_F} \hat s_\theta$. We use $\hat G_F = 1.1663787\times 10^{-5}\, \text{GeV}^{-2}, \hat m_Z = 91.1976\, \text{GeV},$ and $\hat m_W = 80.387\, \text{GeV}$~\cite{Corbett:2021eux}.}

The leading terms are $\mathcal O(\hat s)$ in the cross section, implying amplitudes that grow as $\mathcal O(\hat s)$ as well.\footnote{The subleading terms are: $
\hat \sigma_{\hat s \to\infty}(\bar q q' \to \ell^{\pm}\nu)_{\hat s \to\infty, \mathcal O(s^0)}  = \frac{\hat s\, x\, \hat e^2}{72\,\pi\, \hat v^2\, \hat s^2_\theta}\, C^{3,(6)}_{LQ} + \frac{x^2\, \hat e^2}{576\, \pi\, \hat v^4\, \hat s^2_\theta}\Big( 8\, \hat v^2\, C^{3, (6)}_{LQ}\, (C^{3,(6)}_{HL} + C^{3,(6)}_{HQ} - 2\,\sqrt 2\,\delta G^{(6)}_F) + 4\, \hat v^2\,C^{2,(8)}_{LQ} + 2\,\hat m^2_W(C^{t,3,(8)}_{LQ} - 4\, C^{s,3,(8)}_{LQ}) \Big)$}  As expected from power counting, the operators that show up in the large $\hat s$ limit are dimension six contact terms squared and the dimension eight contact terms involving derivatives. The dimension eight terms enter proportional to the SM couplings as they interfere with the SM, while the dimension six squared terms have no factors. As such, if all Wilson coefficients have roughly the same size, one expects that the SM factors will suppress the dimension eight pieces relative to dimension six squared\footnote{Following the arguments in Ref.~\cite{Giudice:2016yja} based on $\hbar$ counting, matching to dimension six, four fermi operators generates coefficients with the same UV coupling order as matching to dimension eight four fermi operators with two derivatives. So, dimension six squared terms carry additional powers of UV coupling compared to the interference of their dimension eight, two derivative, counterparts with the SM (see Ref.~\cite{Hays:2020scx} for an example). The SM couplings accompanying the interference piece replace the `missing' UV coupling. In this light, the assumption that dimension six and eight coefficients are the same size can be restated as the assumption that UV couplings are $\mathcal O(1)$.}.  A further difference between the different $\mathcal O(1/\Lambda^4)$ effects is the sign.  The dimension six squared piece is positive definite, while the sign of the dimension eight piece depends on the sign of the Wilson coefficients. Depending on their sign, they can cancel each other or conspire to cancel the dimension six squared piece. The dependence on sign and role of interference seen here echoes what was seen in Ref.~\cite{Corbett:2021cil,Martin:2021vwf} for SM loop induced processes involving Higgses.

Finally, an important aspect of $W$ production is that only one dimension six contact operator (with the flavor assumptions we have made) -- $C^{3,(6)}_{LQ}$ -- enters. For a positive value of this coefficient -- the choice made so far -- both the interference of $C^{3,(6)}_{LQ} \times SM$ and the squared term are positive, so there is no potential cancellation between the dimension six linear and quadratic pieces; such cancellations can lead to more erratic behavior in the differential cross section. With no possible cancellations or oddities in the dimension six piece, studies of the impact of higher order $\mathcal O(1/\Lambda^4)$ terms are easier to carry out. This situation can be contrasted with dilepton production, where there are multiple dimension six contact terms and the competing terms don't all have the same sign interference with the SM. As a consequence, concrete conclusions about $\mathcal O(1/\Lambda^4)$ effects are harder to draw, as they can change dramatically with a sign flip in the dimension six coefficients. For more details, see Appendix~\ref{sec:dilepton}.

\section{Proton level results}
\label{sec:results}

To generate proton level results, we use the NNPDF3.0 NLO parton distribution functions~\cite{Hartland_2013, Ball_2015} for $\alpha_s = 0.118$ and factorization and renormalization scales set by default to $m^2_Z$. We take the LHC energy to be $\sqrt s = 13\, \text{TeV}$.

 In order to better illustrate the SMEFT effects, and the effects from $\mathcal O(1/\Lambda^4)$ in particular, we will assume numerical values for the Wilson coefficients throughout this section. Clearly, our assumptions about the coefficients (sign, any hierarchy among them) will strongly influence our results, so the reader must bear this in mind. We will comment on any specific coefficient signs/hierarchies that have particularly strong effects, and the complete $\mathcal O(1/\Lambda^4)$ expressions for any other coefficient assumptions can be generated using the expressions in Sec. \ref{sec:thesetup}.

\begin{figure}[h!]
\includegraphics[width=0.49\textwidth]{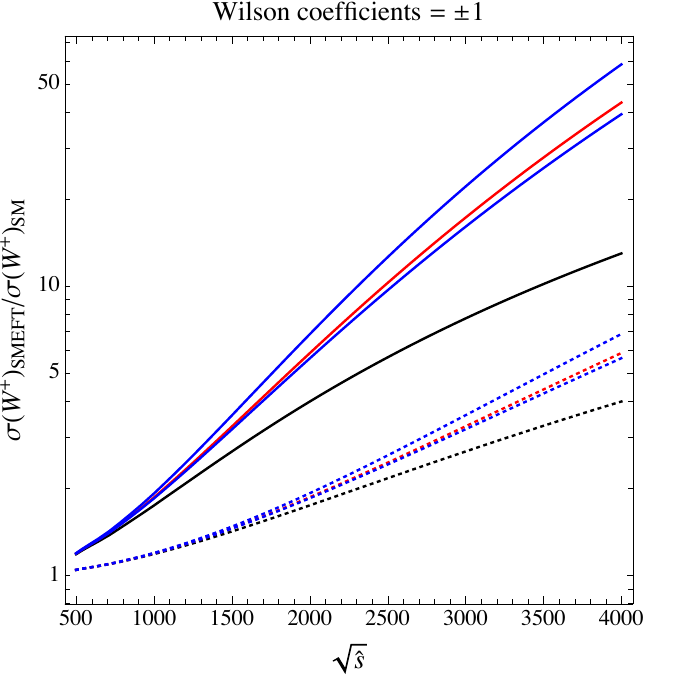}
\includegraphics[width=0.49\textwidth]{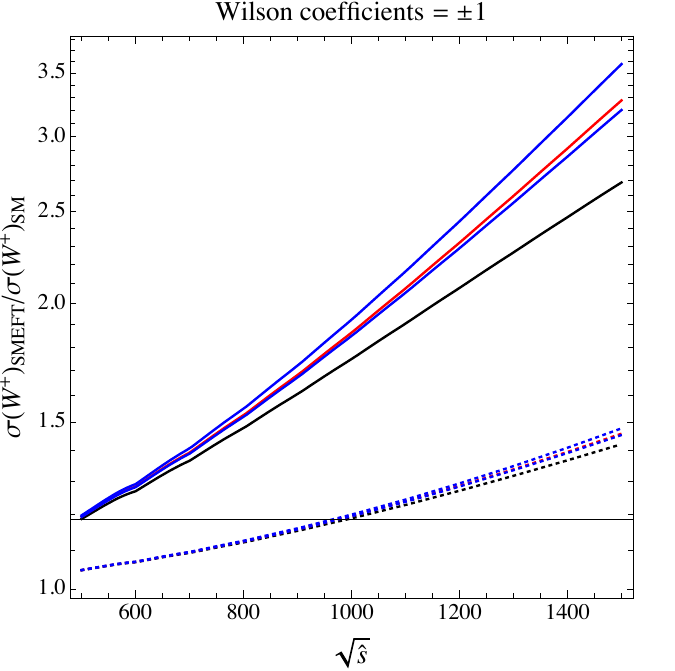}
\caption{Differential cross section $d\sigma(p p \to \ell^+\nu)/d\sqrt{\hat s}$ calculated in SMEFT compared to the SM value as a function of $\sqrt{\hat s}$, with all Wilson coefficients set to a magnitude of 1 and a new physics scale of $5\, \text{TeV}$ for the solid curves and $10\, \text{TeV}$ for the dotted curves. The black lines show the ratio with the SMEFT calculation truncated at $\mathcal O(x) = \mathcal O(1/\Lambda^2)$ while the red and blue curves show the ratio using the full $\mathcal O(x^2)$ SMEFT result. The red line shows the effect when we take all dimension eight coefficients to be positive, and the blue lines enveloping the red line show the result where we choose the signs of $C^{s,3,(8)}_{LQ}$ and $C^{t,3,(8)}_{LQ}$ to maximize/minimize the leading $\hat s$ behavior shown in Eq.~\eqref{eq:leadings}. The right panel is just a zoomed in version of the left panel. Changing the factorization scheme from $\mu^2_F = m^2_Z$ to $\mu^2_F = \hat s$ has no effect on these ratios. }
\label{fig:dsigmadm}
\end{figure}

We proceed with the assumption that all Wilson coefficients have a value of 1, so that the only variable in our expressions is $x$, which we can exchange for the new physics scale $\Lambda$. In addition to being simple, this coefficient size is backed up by the fact that contact terms and the majority of the coefficients entering the coupling and width expansion in Eq.~\eqref{eq:coeff1}-\eqref{eq:coeff4} are generated at tree-level by generic weakly coupled UV physics, following the classification scheme developed in Ref.~\cite{Arzt:1994gp,Craig:2019wmo}. With the coefficients set, we explore how physics with fixed $\Lambda$ affects the differential cross section $d\sigma(p p \to \ell^+\nu)/d\sqrt{\hat s}$. In Fig.~\ref{fig:dsigmadm}, we show the ratio of the SMEFT result to $\mathcal O(1/\Lambda^2)$ and $\mathcal O(1/\Lambda^4)$ to the SM result. For the $\mathcal O(1/\Lambda^4)$ result we show three different curves, one with all Wilson coefficients equal to $+1$, one where the sign of $C^{t,3,(8)}_{LQ}$ is flipped to maximize the dimension eight part of Eq~\eqref{eq:leadings}, and one where the sign of $C^{s,3,(8)}_{LQ}$ is flipped to minimize the effect. We do this in an attempt to bound the region of dimension eight effects, at least for situations where all the Wilson coefficients are roughly the same size.

A primary motivation for calculating $\sigma(pp \to \ell^+\nu)$ at $\mathcal O(1/\Lambda^4)$ is to use it as a laboratory for higher order effects, meaning to systematically study how the hierarchy of higher order terms behaves in different kinematic regimes and under different UV assumptions. Using this data, and data from other processes, we hope to better inform truncation error estimates at more complex LHC processes where full $\mathcal O(1/\Lambda^4)$ results are not available. See Ref.~\cite{Trott:2021vqa} for more on this philosophy.  As a first example, we consider $W^+$ production between a minimum center of mass energy $\sqrt{\hat s_{min}}$ and maximum $\sqrt{\hat s_{max}}$, where $\hat s$ is the invariant mass squared of the final state particles. Technically, for leptonic $W^+$, we should form the transverse mass rather than the invariant mass, but we will work with the invariant mass for this exercise as it is simpler. We would like to know the role of the $\mathcal O(1/\Lambda^4)$ pieces in this analysis, and what role the dimension eight operators play compared to dimension six squared. 

\begin{figure}[h!]
\includegraphics[width=0.49\textwidth]{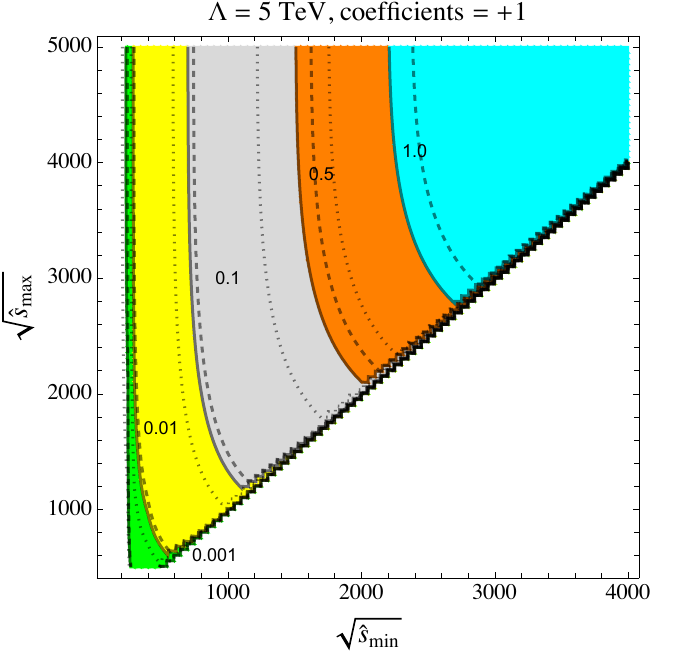}
\includegraphics[width=0.49\textwidth]{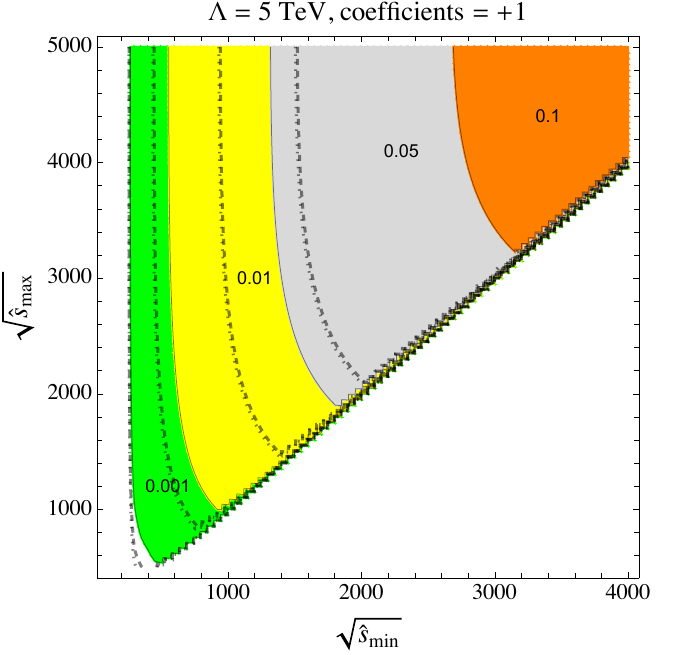}
\caption{In the left panel, we show the absolute value of the ratio of the cross section in the full, $\mathcal O(1/\Lambda^4)$ calculation to the linear $\mathcal O(1/\Lambda^2)$ calculation as a function of the minimum center of mass energy $\sqrt{\hat s_{min}}$ and the maximum $\sqrt{\hat s_{max}}$: $|(\sigma(pp \to \ell^+\nu)_{\mathcal O(x^2)} - \sigma(pp \to \ell^+\nu)_{\mathcal O(x)} )/\sigma(pp \to \ell^+\nu)_{\mathcal O(x)}|$. All Wilson coefficients are taken to be positive and equal to one, and the new physics scale $\Lambda = 5\,\text{TeV}$. In the right plot we show the ratio (absolute value) of the $\mathcal O(1/\Lambda^4)$ calculation to the cross section including only dimension six operators as a function of  $\sqrt{\hat s_{min}}$ and $\sqrt{\hat s_{max}}$: $|(\sigma(pp \to \ell^+\nu)_{\mathcal O(x^2)} - \sigma(pp \to \ell^+\nu)_{\mathcal O(x^2), no\, dim-8} )/\sigma(pp \to \ell^+\nu)_{\mathcal O(x^2), no\ dim-8}|$. The lower right portion is unphysical as $\sqrt{\hat s_{min}} > \sqrt{\hat s_{max}}$, and the step-like structure of the edge of this boundary is a numerical artifact. The dotted (dashed) black lines represent the contours where we have chosen the signs of $C^{s,3,(8)}_{LQ}$ and $C^{t,3,(8)}_{LQ}$ to maximize (minimize) Eq.~\eqref{eq:leadings} instead of taking all dimension eight coefficients to be $+1$. }
\label{fig:ratiosWp}
\end{figure}

To test the role of $\mathcal O(1/\Lambda^4)$, we calculate the ratio of the integrated cross section between $\sqrt{\hat s_{min}}$ and maximum $\sqrt{\hat s_{max}}$ including all $\mathcal O(1/\Lambda^4)$ effects compared to the integrated cross section with the same endpoints but only including the linear, $\mathcal O(1/\Lambda^2)$ effects. To get a numerical value, we need to make an assumption for the Wilson coefficient sizes and the size of $x = \bar v^2_T/\Lambda^2$. Taking all coefficients to 1 and setting $\Lambda = 5\, \text{TeV}$, the ratio is shown below in the left panel Fig.~\ref{fig:ratiosWp} as a function of the minimum and maximum center of mass energy.
In the right panel we show a similar calculation, except that the denominator of the ratio is the integrated cross section including dimension six linear {\em and} squared terms  -- $\sigma(pp \to \ell^+\nu)_{\mathcal O(x^2),no\ dim-8}$-- thereby showing the effect of dimension eight operators alone. In the dashed and dotted lines of Fig~\ref{fig:ratiosWp}, we repeat the calculation after picking the signs of the dimension eight operators to maximize (dotted) or minimize (dashed) their effect according to Eq.~\eqref{eq:leadings}. For both plots we begin at $\sqrt{\hat s_{min}} = 100\, \text{GeV}, \sqrt{\hat s_{max}} = 500\, \text{GeV}$.

Examining Fig~\ref{fig:ratiosWp}, the impact of the $1/\Lambda^4$ terms increases as we increase $\sqrt{\hat s_{min}}$. Picking an  $\sqrt{\hat s_{min}}$ and varying  $\sqrt{\hat s_{max}}$ the results quickly asymptote, indicating that, at least for this choice of Wilson coefficient and scale $\Lambda$, the suppression of the parton distribution functions at larger energy overwhelms any growth in energy in the partonic cross section, so the net result is dominated by physics at $\sqrt{\hat s_{min}}$. This matches the behavior found in Ref.~\cite{Cohen:2021gdw}.  Focusing on the left plot, we see that -- at least for the choice of all Wilson coefficient $\mathcal O(1)$ -- the impact of the $\mathcal O(1/\Lambda^4)$ terms is $\mathcal O(10\%)$ when $\sqrt{\hat s_{min}} \sim 1\, \text{TeV}$, but grows to $\mathcal O(50\%)$ when $\sqrt{\hat s_{min}} \sim 2\, \text{TeV}$. For these $\sqrt{\hat s_{min}}$ values, the right hand plot indicates the dimension eight operators are only a small fraction, $\mathcal O(\text{few}\%)$ of the $\mathcal O(1/\Lambda^4)$ result (again for this coefficient choice), with the brunt of the effect coming from dimension six squared terms\footnote{If we replace $\sqrt{\hat s}$ in these plots with $m_T$ -- a more physical variable for leptonic $W$ production -- the qualitative picture is unchanged. Had we chosen a negative value for the dimension six, the $\mathcal O(1/\Lambda^2)$ interference is negative while the $\mathcal O(1/\Lambda^4)$ term is positive. In this case, where cancellation between different order terms is possible, the impact of higher dimensional SMEFT operators is inflated.}.

In the left panel, the dashed and dotted contours straddle the original contour, with the $\sqrt{\hat s_{min}}$ separation between the dashed and dotted increasing as $\sqrt{\hat s_{min}}$ increases. In the right hand plot, the dashed and dotted contours nearly overlap and are offset from the original contour (remember these are the absolute value of the cross section ratios). For example, the dashed/dotted contours for ratio of $0.05$ sit at $\sqrt{\hat s_{min}} \sim 1\, \text{TeV}$, right in the middle of the $0.01$ contour when all signs are positive.  The fact that the min/max dimension eight contours do not envelop the contour with all coefficients taken to be positive may seem confusing at first, however the mix/max signs were chosen by examining only the $\mathcal O(\hat s)$ piece of the partonic cross section in Eq.~\eqref{eq:leadings}. So, the fact that the contours do not overlap is just a sign that the $\mathcal O(\hat s^0)$ piece is also important when calculating the ratio in the right hand plot and that it has a different sign dependency than the $\mathcal O(\hat s)$ piece.

In the above test, we took all Wilson coefficients to have the same size. While a reasonable first guess, it does not capture the full range of effects from higher order terms. To explore how overall suppression or enhancement of the dimension eight versus dimension six coefficients affects things, we focus on two particular ranges of center of mass energy $1\, \text{TeV} \le \sqrt{\hat s} \le 2\, \text{TeV}$ and $2\, \text{TeV} \le \sqrt{\hat s} \le 3\, \text{TeV}$. In these two energy bins, we set all coefficients for a given mass dimension to a common value -- $C^{(6)}$ for all dimension six coefficients and $C^{(8)}$ for all dimension eight.  Then, we calculate the size of the dimension eight piece compared to the rest of the $\mathcal O(1/\Lambda^4)$ result as a function of the common dimension six coefficient and the ratio of the dimension eight coefficient to the dimension six coefficient ($C^{(8)}/C^{(6)}$). The result, shown in Fig.~\ref{fig:varycoefficients} below, is the identical calculation to the right panel of Fig.~\ref{fig:ratiosWp} except we are focusing on two representative $\sqrt{\hat s_{min}}, \sqrt{\hat s_{max}}$ values and allowing the relative size of different order coefficients to vary.
\begin{figure}[h!]
\includegraphics[width=0.49\textwidth]{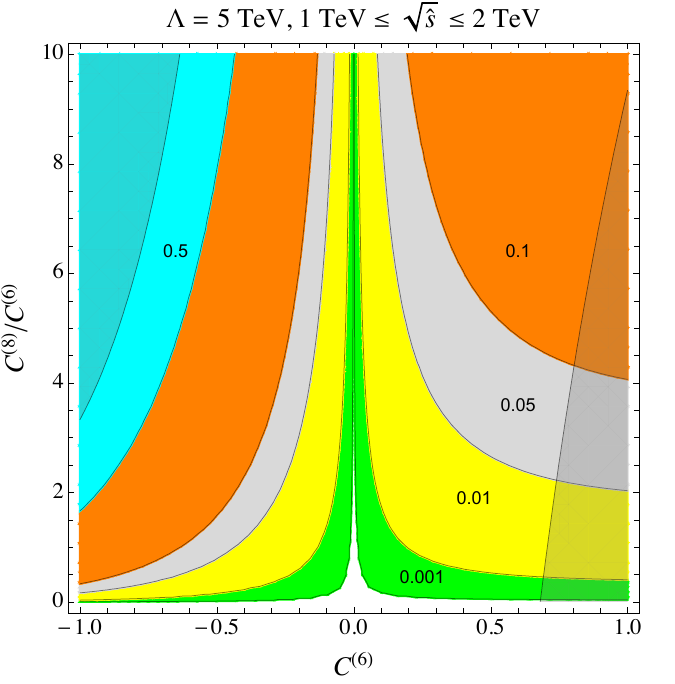}
\includegraphics[width=0.49\textwidth]{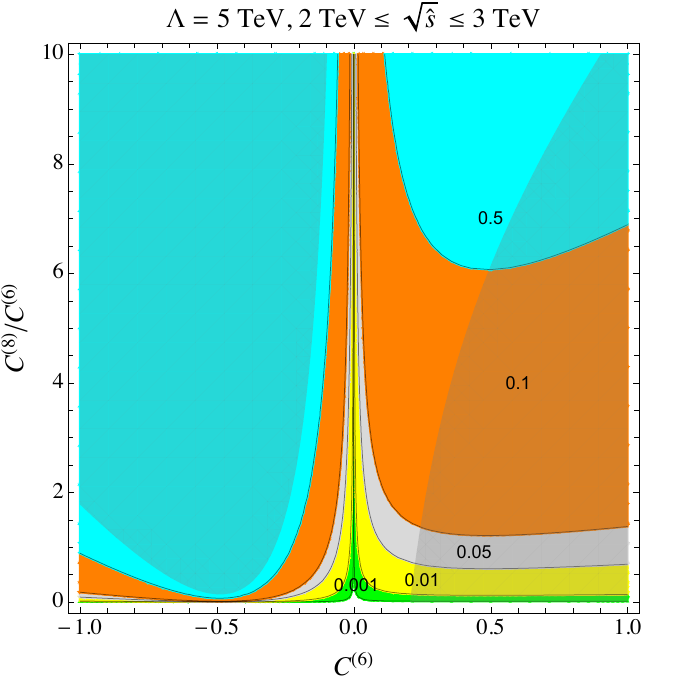}
\caption{Dimension eight operator contribution to $\sigma(p p \to \ell^+\nu)$ in the center of mass energy range $1\, \text{TeV} \le \sqrt{\hat s} \le 2\, \text{TeV}$ (left panel)  and $2\, \text{TeV} \le \sqrt{\hat s} \le 3\, \text{TeV}$ (right panel) compared to the full $\mathcal O(1/\Lambda^4)$ result with the dimension eight coefficients shut off: $|(\sigma(pp \to \ell^+\nu)_{\mathcal O(1/\Lambda^4)} - \sigma(pp \to \ell^+\nu)_{\text{no\, dim-8}})/\sigma(pp \to \ell^+\nu)_{\mathcal O(1/\Lambda^4)}|$ plotted as a function of the dimension six Wilson coefficient strength $C^{(6)} = [-1.0, 1.0]$ and the ratio of the dimension eight coefficient strength relative to the dimension six coefficient (taking all dimension eight coefficients to be equal), $C^{(8)}/C^{(6)} = [0.1,10]$. The new physics scale $\Lambda = 5\, \text{TeV}$ in both panels. The shaded region indicates where the SMEFT contribution is larger than the SM contribution (either positive or negative).}
\label{fig:varycoefficients}
\end{figure}

For fixed $C^{(6)}$, the denominator of the ratio plotted is fixed. Increasing $C^{(8)}/C^{(6)}$, we see that the impact of dimension eight grows. We have included both positive and negative values for $C^{(6)}$ but taken the dimension eight coefficients to be positive for simplicity\footnote{One motivation for taking positive dimension eight coefficients is the conditions analyticity can impose~\cite{Adams:2006sv, 2014}. However, as pointed out in Ref.~\cite{Li:2022rag}, the bounds are more subtle than just requiring all dimension eight coefficients to be positive.  Extending Fig.~\ref{fig:varycoefficients} to negative values of $C^{(8)}/C^{(6)}$, while still keeping all coefficients equal, the contours mirror $C^{(8)}/C^{(6)} > 0$. Analyticity constraints also can be applied to dimension six coefficients and take the form of sum rules, see~\cite{Gu:2020thj,Azatov:2021ygj, Remmen:2020uze, Bellazzini:2020cot}. We have not imposed constraints from these on our current analysis, though it would be interesting to do so.}. The shaded regions on the plot show where the SMEFT contribution to the cross section in this kinematic region is the same size as the SM piece, included to help orient the contours in terms of actual observables, and they can be viewed as rough bounds\footnote{The contours are essentially unchanged if we neglect all dimension six coefficients other than $C^{3,(6)}_{LQ}$ or if we plot in bins of $m_T$ instead of $\sqrt{\hat s}$.}.  More accurate bounds require correctly incorporating the appropriate experimental acceptance/efficiencies and are left for future work.

As $C^{(8)}/C^{(6)}$ is varied from $0.1$ to $10$, the effect of the dimension eight terms increases by roughly two orders of magnitude. For example, fixing  $C^{(6)} = 0.1$ and varying $C^{(8)}/C^{(6)}$ the impact of the dimension eight piece varies grows from $5.7 \times 10^{-4}$ to $0.057$ (for $\sqrt{\hat s} \in [1\, \text{TeV}, 2\, \text{TeV}]$). The effects of $C^{(8)}/C^{(6)}$ are larger for negative $C^{(6)}$ because of a cancellation between the negative $\mathcal O(1/\Lambda^2)$ interference and positive dimension six squared $\mathcal O(1/\Lambda^4)$ contributions. The overall impact of dimension eight also increases as we move to higher $\hat s$. Note that we can use Fig.~\ref{fig:varycoefficients} to extrapolate the results of Fig.~\ref{fig:ratiosWp} to $\Lambda$ other than $5\, \text{TeV}$, as shifting $\Lambda \to \Lambda'$ is equivalent to rescaling both $C^{(6)}$ and $C^{(8)}/C^{(6)}$ by $(\Lambda/\Lambda')^2$.

Had we calculated the net $\mathcal O(1/\Lambda^4)$ relative to the $\mathcal O(1/\Lambda^2)$ result -- analogous to the left plot of Fig.~\ref{fig:ratiosWp} -- for the same inputs, the result would depend more sensitively on the individual coefficient sign choice. However, for $\sqrt{\hat s} \in [1\, \text{TeV}, 2\, \text{TeV}]$ this ratio is driven by the dimension six coefficient, to the extent that the ratio when $C^{(6)} = 0.1, C^{(8)}/C^{(6)} = 10$ is smaller than when both coefficients are 1. This remains true at higher $\hat s$. 

Summarizing, we have found that $\mathcal O(1/\Lambda^4)$ terms have an $\mathcal O(10\%)$ impact on the cross section at $\sqrt{\hat s} = 1\, \text{TeV}$ for new physics scale of $\Lambda = 5\, \text{TeV}$ and all Wilson coefficients $\mathcal O(1)$. The effect grows as we look at higher energies and is $\sim 50\%$ at $\sqrt{\hat s} = 2\, \text{TeV}$. We have focused on the $1 - 2\, \text{TeV}$ range as it is far from the $W$ resonance region yet where the cross section is still high enough that the LHC should have significant amounts of data. For this choice of $\Lambda$, coefficient hierarchy, and kinematic region, the  overall $\mathcal O(1/\Lambda^4)$ effect is driven by the dimension six squared piece and largely insensitive to the signs of various dimension eight operators. This is due, at least in part, to the fact that only one dimension six operator appears in the large $\hat{s}$ limit, so no cancellations among different dimension six terms can occur. The exact contribution to $\mathcal O(1/\Lambda^4)$ coming from dimension eight is more sensitive to sign choices, but lies in the few percent range for our $\Lambda$ choice. Breaking the assumption that all Wilson coefficients are $\mathcal O(1)$, we get a wider range of effects.  Shrinking the dimension six coefficients relative to dimension eight for fixed $\Lambda$, the net $\mathcal O(1/\Lambda^4)$ effect goes down, but a larger portion of it comes from dimension eight.

Our study is by no means exhaustive but does give us an estimate of the $\mathcal O(1/\Lambda^4)$ contribution and how it varies with model/Wilson coefficient assumptions. We emphasize that we have only plugged in numbers here to be quantitative. Our expressions in Sec.~\ref{sec:thesetup} are valid independent of any specific UV assumptions other than the usual SMEFT assumptions (no light degrees of freedom other than the SM, and only one source of electroweak symmetry breaking) and our flavor assumption.

\section{Four fermion contributions to $pp \to \ell^+\ell^-, \ell^{\pm}\nu$ at arbitrary mass dimension}
\label{sec:allorders}
As discussed earlier, four fermion operators lie outside of the geoSMEFT scope. Specifically, as the kinematics of $4^+$ point vertices is no longer trivial, we can always generate new operators by attaching more derivatives. While this means the number of operators no longer plateaus with increasing mass dimension, we can still find a relatively compact form. 

The first important thing to notice is that, in order to contribute to the processes of interest, operators must have the form $D^{n}H^m\psi^4$. No operators with field strengths can contribute at tree level\footnote{Contact operators of the form $GGLL, GGe_ce_c$ etc. can contribute to $pp \to \ell^+ \ell^-$. However they do not interfere with the SM contribution and must enter as (contact term)$^2$.}. Next, if any derivatives act on a Higgs, then the operator either has a physical Higgs boson in it or a (longitudinal) gauge boson. In either case, the operator field content is not correct to contribute to $pp \to \ell^+\ell^-, \ell^{\pm}\nu$ (at tree level). Therefore, we're really only looking for operators of the form $H^m D^n(\psi^4)$, where setting the Higgs to its vev leaves us with an operator with only four fermions.

In order to interfere with the SM $\bar q q' \to \ell^+\ell^-, \ell^{\pm}\nu$, the fermions in $\psi^4$ must have the form $\psi^{\dag}_1\psi_1 \psi^\dag_2 \psi_2$ where $\psi_1 = Q, u_c, d_c$ and $\psi_2 = L, e_c$. Operators that don't interfere with the SM have less restricted fermion form, but they will always enter at a higher power of $\Lambda$ than operators that can interfere. Focusing on operators that interfere, the subset of operators we care about are $(H^\dag H)^m D^{2n}(\psi^{\dag}_1\psi_1 \psi^\dag_2 \psi_2)$. The fact that only even powers of derivatives and Higgses can appear is set by the fermionic structure.

In this form and with no identical fields present, the operator is factorized -- the EW structure of $(\psi^{\dag}_1\psi_1 \psi^\dag_2 \psi_2)$ determines the EW structure of the Higgs piece and is completely independent of how the derivatives are sprinkled among the four fermions. The number of EW structures depends on the helicity of the fermions involved. For $\psi_1 = Q, \psi_2 = L$, there are 6 possible structures, for $\psi_1 = u_c, d_c, \psi_2 = L$ or $\psi_1 = Q, \psi_2 = e_c$ there are 2, while for $\psi_1 = u_c, d_c, \psi_2 = e_c$ there is only 1. The EW forms are identical to those shown in Eq.~\eqref{eq:contactH1}, each multiplied by $(H^{\dag}H)^{m-1}$, supplemented by an additional structure for $\psi_1 = Q, \psi_2 = L$ that only shows up at dimension $\ge 10$, 
\begin{align}
(H^{\dag}H)^{m-2}((H^{\dag}\tau^I H)(H^{\dag} \tau^J H) + I \leftrightarrow J)(Q^\dag \bar{\sigma}^\mu \tau_I Q)(L^{\dag} \bar{\sigma}_\mu\,\tau_J L) \nonumber
\end{align}

For each EW structure, we spread the derivatives among the fermions, $D^{2n}(\psi^{\dag}_1\psi_1 \psi^\dag_2 \psi_2)$. As in Sec~\ref{sec:opsdim8}, this is best done in momentum space. As we saw there, for four particles there are only two invariants we can form after accounting for the equations of motion ($p^2_i = 0$ for massless fermions) and momentum conservation (IBP). The two are a subset of the Mandelstam $s, t, u$, where the condition $s + t + u = 0$ can be used to remove one invariant. With $D^2$, and choosing to remove $u$, the only options were terms linear in $s$ or $t$. Generalizing this to $2n$ derivatives, all terms of the form $s^m t^{n-m}$ are possible,
\begin{align}
D^{2n}(\psi^{\dag}_1\psi_1 \psi^\dag_2 \psi_2) \to \sum\limits_{m=0}^n\, s^m\, t^{n-m}(\psi^{\dag}_1\psi_1 \psi^\dag_2 \psi_2). 
\end{align}
For example, at $D^4$ there are three possibilities for a given EW structure:
\begin{align}
(C^{(2,0)}_{\psi^4}\, s^2 + C^{(1,1)}_{\psi^4}\, s\, t + C^{(0,2)}_{\psi^4}\, t^2 )(\psi^{\dag}_1\psi_1 \psi^\dag_2 \psi_2). 
\end{align}
This form persists to arbitrary mass dimension, and is generated by
\begin{align}
\frac{1}{(1-s)(1-t)}(\psi^{\dag}_1\psi_1 \psi^\dag_2 \psi_2), 
\label{eq:momgen}
\end{align}
as can be shown via Hilbert series techniques~\cite{Lehman:2015via, Lehman:2015coa, Henning:2015daa, Henning:2015alf, Henning:2017fpj}, or by directly finding the quotient ring for four momenta subject to overall momentum conservation and the EOM. While expanding Eq.~\eqref{eq:momgen} to $\mathcal O(s^m t^{n-m})$ gets us the number of operators with $2n$ derivatives and their dependence on $s,t$, each of those operators, in principle, carries an independent coefficient, and once the coefficients are included the net result cannot be resummed (for generic values of the coefficients).

Putting the pieces together, the $(H^\dag H)^m D^{2n}(\psi^{\dag}_1\psi_1 \psi^\dag_2 \psi_2)$ operators can be written as
\begin{align}
\sum\limits_{k=0}^n\, \frac{C^{(i),(k,n-k)}_{\psi^4}}{\Lambda^{2(m+n+1)}}\,s^k\, t^{n-k} (H^\dag H)_i^m  (\psi^{\dag}_1\psi_1 \psi^\dag_2 \psi_2)_i
\end{align}
where $i$ labels the EW structure and it is understood that $s,t$ are formed from the $\psi_i$ momenta only. For fixed operator dimension, the operators with the largest impact on the cross section are those with the strongest momentum dependence. At operator dimension $d$, the maximal momentum power that can accompany the operator is $d-6$.

Putting this to use, we can estimate the size of $\mathcal O(1/\Lambda^6)$ terms in different kinematic regions. The most dangerous terms at $\mathcal O(1/\Lambda^6)$ come from dimension ten four fermion operators with four derivatives, and the interference between dimension six and two-derivative, dimension eight  four fermion terms. Focusing on this subset\footnote{At $\mathcal O(1/\Lambda^6)$ there are also corrections to the subleading terms (in the large $\hat s$ limit) in the amplitude, but these corrections are suppressed by a factor of $x = \bar v^2_T/\Lambda^2$, i.e. $\hat s\,  \bar v^2_T/\Lambda^2$. We have also only kept terms consistent with flavor $(U(3))^5$.}, assuming a single electroweak structure (the four derivative analog of $C^{s/t,3,(8)}_{LQ}$) for the dimension ten operators, and taking all Wilson coefficients to be equal to one, we have
\begin{align}
& |A^W_{LL, \mathcal  O(x^3), C_i = 1} |^2 \sim 4\, \hat u^2\ \Big( \frac{g^0_{W, f_L}g^0_{W,\ell_L}}{\hat s - m^2_W} + \frac{x}{v^2_T} - \frac{x^2\ (\hat s + \hat t)}{v^4_T}  +  \frac{x^3\, (\hat s^2 + \hat t^2 + \hat s\,\hat t)}{v^6_T} \Big)^2\Big|_{\mathcal O(x^3)} \rightarrow \nn
& ~~~~~~~~~~~~~~~~~~~~~~ \hat \sigma_{\mathcal O(x^3), C_i = 1} \sim -\frac{x^3\, s^2}{2880\,\pi\, \hat v^6}(30 - 17\, \hat e^2/\hat s^2_\theta)
\end{align}
Using this, we can explore the ratio of the $\mathcal O(1/\Lambda^6)$ contribution to the full $\mathcal O(1/\Lambda^4)$ result -- the analog of Fig.~\ref{fig:ratiosWp} but with $\mathcal O(1/\Lambda^6)$ in the numerator and $\mathcal O(1/\Lambda^4)$ in the denominator. The result is shown below in Fig.~\ref{fig:ratiodim10} as a function of the minimum and maximum center of mass energy of the process, and as in Fig.~\ref{fig:ratiosWp} we have taken the new physics scale $\Lambda = 5\,\text{TeV}$. 

\begin{figure}[h!]
\includegraphics[width=0.49\textwidth]{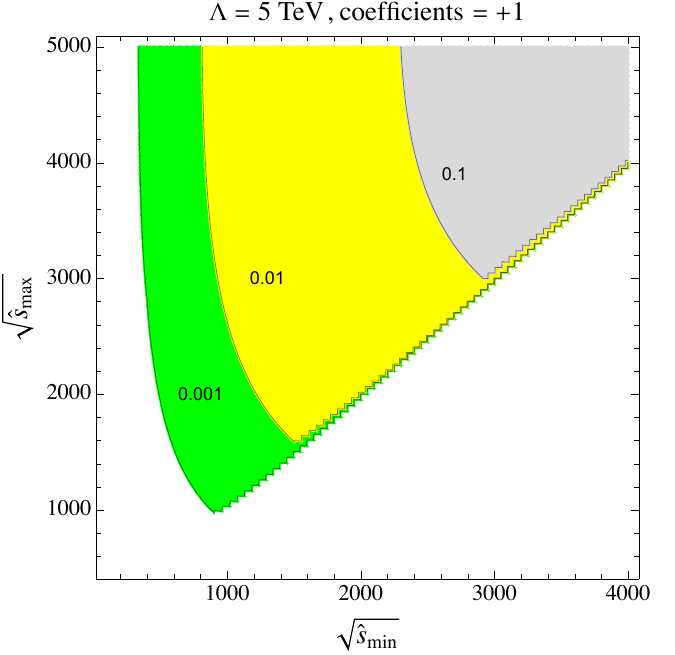}
\caption{Absolute value of the ratio of the $\mathcal O(1/\Lambda^6)$ contributions relative to the SMEFT cross section to $\mathcal O(1/\Lambda^4)$ as a function of the minimum center of mass energy $\sqrt{\hat s_{min}}$ and the maximum $\sqrt{\hat s_{max}}$, $|(\sigma(pp \to \ell^+\nu)_{\mathcal O(1/\Lambda^6)\, \text{large\,} \hat s} - \sigma(pp \to \ell^+\nu)_{\mathcal O(1/\Lambda^4)}) /\sigma(pp \to \ell^+\nu)_{\mathcal O(1/\Lambda^4)}|$). All Wilson coefficients are taken to be positive and equal to one, and the new physics scale $\Lambda = 5\,\text{TeV}$.}
\label{fig:ratiodim10}
\end{figure}

Comparing Fig~\ref{fig:ratiodim10} with Fig.~\ref{fig:ratiosWp}, we see that the $\mathcal O(1/\Lambda^6)$ effects are one to two orders of magnitude smaller than the $\mathcal O(1/\Lambda^4)$ effects for the kinematic regime we have plotted.

\section{Conclusions}
\label{sec:conclude}

In this paper we have studied SMEFT corrections to monolepton production at $\mathcal O(1/\Lambda^4)$, adding it to the list processes~\cite{Hays:2018zze,Hays:2020scx, 2020,Corbett:2021eux,Corbett:2021iob} now known to that order. We are particularly interested in the relative size of the $\mathcal O(1/\Lambda^2)$ and $\mathcal O(1/\Lambda^4)$ pieces and how that changes as we push into the $\hat s \gg m^2_W$ regime. The main goal of this study is to extend the list of processes known (with flavor assumptions), joining Ref.~\cite{Hays:2018zze,Hays:2020scx, 2020, Corbett:2021eux}, up to $\mathcal O(1/\Lambda^4)$, with the hope that a broader index of exact results will improve how truncation uncertainties in EFT analyses are estimated.

To keep the calculation manageable and focus on the large $\hat s$ effects, we have assumed $(U(3))^5$ symmetry with massless fermions and diagonal CKM and PMNS matrices, though the terms with the most severe energy growth are unchanged under a more relaxed assumption of $U(3)_Q\times U(3)_{u+d}\times U(3)_{L}\times U(3)_{e}$. Further relaxing the flavor symmetry has little impact on the dimension eight piece of the calculation, as the flavor structure of those operators are limited by what can interfere with the SM. Dimension six squared terms are free of this restriction, so the number of terms we need to consider grows considerably if we relax the flavor assumption.

While our analytic expressions hold for all coefficient choices and $\Lambda$ values, we performed several numerical studies using the simplification that $\Lambda = 5\,\text{TeV}$. Taking all Wilson coefficients $\pm 1$, we quantified how the importance of $\mathcal O(1/\Lambda^2)$ terms grows with $\sqrt{\hat s_{min}}$, the minimum center of mass energy. Throughout the range of $\sqrt{\hat s_{min}}$ we considered and even allowing for an order of magnitude hierarchy between the dimension six and eight Wilson coefficients, the $\mathcal O(1/\Lambda^4)$ is dominated by the dimension six squared contribution. Part of this stability is due to the fact that, with our flavor assumptions, only one operator at dimension six is present in the large $\hat s$ regime, so no cancellations among terms of the same mass dimension is possible. 

We also demonstrated a closed form for the four fermion operators of arbitrary mass dimension. Unlike 2- and 3-particle vertices, the number does not plateau with increasing mass dimension, as we can always add more derivatives, however the number and pattern of the higher derivative terms is predictable. Using this form, we can extract the dimension ten operators with the largest energy growth and estimate the effect of $\mathcal O(1/\Lambda^6)$ SMEFT terms as a function of the kinematics. We believe similar all orders expressions are possible for all other four particle (and higher) vertices, though operators with repeated fields and/or multiple Higgs fields will be more complicated as Bose/Fermi statistics will affect the counting of higher derivative terms. It would be interesting to explore how the compact forms for higher point interactions fit into the more geometric picture of effective field theory advocated in Ref.~\cite{Burgess:2010zq, Alonso:2015fsp, Alonso:2016oah,Alonso:2016btr, Helset:2020yio,  Cohen:2020xca,Cohen:2021ucp, Finn:2021jdz, Cheung:2021yog,Cohen:2022uuw,Cheung:2022vnd}.

\acknowledgments
We thank Michael Trott for helpful discussions related to $\hbar$ counting and Chris Hays for comments on the paper. The work of A.M. is partially supported by the National Science Foundation under Grant Number PHY-1820860 and PHY-2112540. 

\appendix
\section{$\sigma(\bar q q \to \ell^{\pm}\nu)$ in the $\hat{\alpha}_{ew}$ input scheme:}
\label{sec:alphascheme}

In the $\hat{\alpha}_{ew}$ scheme, the experimental inputs are the electromagnetic coupling, the $Z$ mass and $G_F$~\cite{Corbett:2021eux}:
\begin{align}
\hat{\alpha}(m_Z) = 1/128.951,\quad \hat m_Z = 91.1976\, \text{GeV},\quad \hat G_F = 1.1663787\times 10^{-5}\, \text{GeV}^{-2} \nonumber 
\end{align}

Using these inputs to fix $g_1, g_2$ and $v_T$, the $\mathcal O(x)$ expansion for the $W$ mass, couplings, and width are:
\begin{align}
& m^0_W = 80.0\, \text{GeV} \nn
& \delta m^{(1)}_W = -62.6\, C^{(6)}_{HWB} - 28.6\, C^{(6)}_{HD} - 24.3\, \delta G^{(6)}_F \nn
& \delta m^{(2)}_W = -31.3\, C^{(8)}_{HWB} -4.29\, C^{(8)}_{HD} - 24.3\, C^{(8)}_{HD,2} - 40.0\, C^{(8)}_{HW,2} - 62.6\, C^{(6)}_{HB}\, C^{(6)}_{HWB} - 124\, (C^{(6)}_{HWB})^2 \nn
& ~~~~~~~~~~~ - 31.7\, C^{(6)}_{HD}\, C^{(6)}_{HWB} - 62.6\, C^{(6)}_{HW}\, C^{(6)}_{HWB} - 134\, C^{(6)}_{HWB}\, \delta G^{(6)}_F + 6.55\, (C^{(6)}_{HD})^2 - 23.6\, C^{(6)}_{HD}\,\delta G^{(6)}_F \nn
& ~~~~~~~~~~~ - 24.7\, (\delta G^{(6)}_F)^2 - 24.3\, \delta G^{(8)}_F 
\end{align}
\begin{align}
& g^0_{W, \ell_L} = -0.46\nn
& \delta g^{(1)}_{W, \ell_L} =  -0.46\, C^{3,(6)}_{HL} + 0.46\, \delta G^{(6)}_F  + 0.16\, C^{(6)}_{HD} + 0.36\, C^{(6)}_{HWB}\nn
& \delta g^{(2)}_{W, \ell_L} =  0.16\, C^{3,(6)}_{HL}\, C^{(6)}_{HD} + 0.36\, C^{3,(6)}_{HL}\, C^{(6)}_{HWB}  + 0.46\, C^{3,(6)}_{HL}\ \delta G^{(6)}_F  + 0.038\, (C^{(6)}_{HD})^2 + \nn
& \quad\quad\quad 0.18\, C^{(6)}_{HD}\, C^{(6)}_{HWB} + 0.019\, C^{(6)}_{HD}\, \delta G^{(6)}_F +0.36\, C^{(6)}_{HW}\, C^{(6)}_{HWB} + 0.7\, (C^{(6)}_{HWB})^2  \nn
& \quad\quad\quad 0.5\, C^{(6)}_{HWB}\, \delta G^{(6)}_F -0.30 (\delta G^{(6)}_F)^2 -0.23\, C^{3,(8)}_{HL} + 0.08\, C^{(8)}_{HD} + 0.08\ C^{(8)}_{HD,2}\nn
& \quad\quad\quad  0.18\, C^{(8)}_{HWB} + 0.23\, i\, C^{(8)}_{\epsilon HL} + 0.46\, \delta G^{(8)}_F\nn
& g^0_{W, \ell_R}  = \delta g^{(1)}_{W, \ell_R}  = \delta g^{(2)}_{W, \ell_R} = 0  
\end{align}
\begin{align}
& g^0_{W, f_L} = -0.46\nn
& \delta g^{(1)}_{W, f_L} =  -0.46\, C^{3,(6)}_{HQ} + 0.46\, \delta G^{(6)}_F  + 0.16\, C^{(6)}_{HD} + 0.36\, C^{(6)}_{HWB}\nn
& \delta g^{(2)}_{W, f_L} =  0.16\, C^{3,(6)}_{HQ}\, C^{(6)}_{HD} + 0.36\, C^{3,(6)}_{HQ}\, C^{(6)}_{HWB}  + 0.46\, C^{3,(6)}_{HQ}\ \delta G^{(6)}_F  + 0.038\, (C^{(6)}_{HD})^2 + \nn
& \quad\quad\quad 0.18\, C^{(6)}_{HD}\, C^{(6)}_{HWB} + 0.019\, C^{(6)}_{HD}\, \delta G^{(6)}_F +0.36\, C^{(6)}_{HW}\, C^{(6)}_{HWB} + 0.7\, (C^{(6)}_{HWB})^2  \nn
& \quad\quad\quad 0.5\, C^{(6)}_{HWB}\, \delta G^{(6)}_F -0.30 (\delta G^{(6)}_F)^2 -0.23\, C^{3,(8)}_{HQ} + 0.08\, C^{(8)}_{HD} + 0.08\ C^{(8)}_{HD,2}\nn
& \quad\quad\quad  0.18\, C^{(8)}_{HWB} + 0.23\, i\, C^{(8)}_{\epsilon HQ} + 0.46\, \delta G^{(8)}_F\nn
& g^0_{W, f_R}  = 0 \nn
& \delta g^{(1)}_{W, f_R}  = \textcolor{blue}{-0.23\, i\, C^{(6)}_{Hud}}\nn
& \delta g^{(2)}_{W, f_R} =  \textcolor{blue}{0.18\, i\, C^{(6)}_{Hud}\, C^{(6)}_{HWB} + 0.08\, i\, C^{(6)}_{Hud}\, C^{(6)}_{HD} + 0.23\, i\, C^{(6)}_{Hud}\, \delta G^{(6)}_F - 0.11\, i\, C^{(8)}_{Hud}}
\end{align}
And for the $W$ width, again for massless fermions and ignoring the CKM matrix:
\begin{align}
& \Gamma^0_W\, (\textrm{GeV}) = 2.01\nn
& \delta\Gamma^{(1)}_W\, (\textrm{GeV}) = 1.34\, C^{3,(6)}_{HL} + 2.68\, C^{3,(6)}_{HQ} - 4.7\, \delta G^{(6)}_F  - 2.16\, C^{(6)}_{HD} - 4.73\, C^{(6)}_{HWB}\nn
& \delta\Gamma^{(2)}_W\, (\textrm{GeV}) =  0.67\, (C^{3,(6)}_{HL})^2 -1.44\, C^{3,(6)}_{HL}\, C^{(6)}_{HD} - 3.15\, C^{3,(6)}_{HL} \, C^{(6)}_{HWB} - 3.1\, C^{3,(6)}_{HL}\,  \delta G^{(6)}_F + 1.3\, (C^{3,(6)}_{HQ})^2 \nn
& \quad\quad\quad-2.9\, C^{3,(6)}_{HQ}\, C^{(6)}_{HD} - 6.3\, C^{3,(6)}_{HL} \, C^{(6)}_{HWB} - 6.2\, C^{3,(6)}_{HL}\,  \delta G^{(6)} - 4.73\, C^{(6)}_{HB}\, C^{(6)}_{HWB} + 1.27\, (C^{(6)}_{HD})^2 \nn
& \quad\quad\quad + 0.99\, C^{(6)}_{HD}\, C^{(6)}_{HWB} + 2.6\, C^{(6)}_{HD}\,  \delta G^{(6)}_F - 4.73\, C^{(6)}_{HW}\, C^{(6)}_{HWB} - 5.6\, (C^{(6)}_{HWB})^2 -0.56\, C^{(6)}_{HWB}\,  \delta G^{(6)}_F \nn
& \quad\quad\quad  + 5.3\, (\delta G^{(6)}_F)^2 + 0.67\, C^{3,(8)}_{HL} + 1.3\, C^{3,(8)}_{HQ} - 0.83\, C^{(8)}_{HD} -1.33\, C^{(8)}_{HD,2} -3.0\, C^{(8)}_{HW,2} - 2.37\, C^{(8)}_{HWB} \nn
& \quad\quad\quad - 4.7\, \delta G^{(8)}_F + \textcolor{blue}{0.34\, (C^{(6)}_{Hud})^2}
\end{align}

\noindent The helicity amplitudes in the $\hat{\alpha}_{ew}$ scheme have extra terms compared to Eq.~\eqref{eq:helampW},\eqref{eq:helampW2} from the expansion of $m_W$ as a series in $x$. The extra terms begin at $\mathcal O(x)$ and only impact the $LL$ helicities (meaning $LR, RL$ and $RR$ expressions match the $\hat m_W$ scheme forms):
\begin{align}
\label{eq:helampW3} 
&\text{Re}(\mathcal A^{W,(1)}_{L,L})_{\hat{\alpha}} = \text{Re}(\mathcal A^{W,(1)}_{L,L})_{\hat m_W} + \frac{2\,\delta m^{(1)}_W\,m_W\, g^0_{W, f_L}\, g^0_{W, \ell_L}\, (\hat s(\hat s-\Gamma^{(0)}_W) + m^4_W - 2 \hat s\, m^2_W)}{P^2_0(\hat s, m_W, \Gamma^{(0)}_W)} \nn
&\text{Im}(\mathcal A^{W,(1)}_{L,L})_{\hat{\alpha}} = \text{Im}(\mathcal A^{W,(1)}_{L,L})_{\hat m_W}+ \frac{\delta m^{(1)}_W\,\Gamma^{(0)}_W\, g^0_{W, f_L}\, g^0_{W, \ell_L}\, (3\, m^4_W + m^2_W((\Gamma^0_W)^2 - 2\,\hat s) - \hat s^2)}{P^2_0(\hat s, m_W, \Gamma^{(0)}_W)} \nn
\end{align}
where, for compactness $m_W$ in these expressions always refers to $m^{(0)}_W$, the zeroth order piece. Similarly, at $\mathcal O(x^2)$:
\begin{align}
& \text{Re}(\mathcal A^{W,(2)}_{L,L})_{\hat{\alpha}} =\text{Re}(\mathcal A^{W,(2)}_{L,L})_{\hat m_W} + \frac{2\,\delta m^{(1)}_W\, m_W(m^4_W - 2\,m^2_W\, \hat s + \hat s(\hat s - (\Gamma^0_W)^2)(g^0_{W, f_L}\,\delta g^{(1)}_{W, \ell_{L}} + g^0_{W, \ell_L}\,\delta g^{(1)}_{W, f_{L}})}{P^2_0(\hat s, m_W, \Gamma^{(0)}_W)}  \nn
&~~+ \frac{(\delta m^{(1)}_W)^2\, g^0_{W, f_L}\,g^0_{W, \ell_L}(3\,(\Gamma^{(0)}_W)^4\, m^2_W\, \hat s - (m^2_W - \hat s)^3(3\, m^2_W + \hat s) + (\Gamma^{(0}_W)^2(m^2_W - \hat s)(m^4_W + 10\, m^2_W\, \hat s + \hat s^2))}{P^3_0(\hat s, m_W, \Gamma^{(0)}_W)}  \nn
& ~~+ \frac{2\,\delta m^{(2)}_W\,m_W\, g^0_{W, f_L}\, g^0_{W, \ell_L}\, ((m^2_W - \hat s)^2 - (\Gamma^{(0)}_W)^2\,\hat s)}{P^2_0(\hat s, m_W, \Gamma^{(0)}_W)} \nn 
& ~~ - \frac{4\,\delta m^{(1)}_W\,\delta \Gamma^{(1)}_W\, m_W\, \Gamma^0_W\,g^0_{W, f_L}\, g^0_{W, \ell_L} (2\, m^6_W - 3\,m^4_W\,\hat s - (\Gamma^0_W)^2\, m^2_W\, \hat s + \hat s^3)}{P^3_0(\hat s, m_W, \Gamma^{(0)}_W)}
\end{align}

\begin{align}
& \text{Im}(\mathcal A^{W,(2)}_{L,L})_{\hat{\alpha}} =\text{Im}(\mathcal A^{W,(2)}_{L,L})_{\hat m_W} + \frac{\delta m^{(1)}_W\,\Gamma^0_W\, (3\, m^4_W + m^2_W((\Gamma^0_W)^2 - 2\hat s) - \hat s^2)(g^0_{W, f_L}\,\delta g^{(1)}_{W, \ell_{L}} + g^0_{W, \ell_L}\,\delta g^{(1)}_{W, f_{L}})}{P^2_0(\hat s, m_W, \Gamma^{(0)}_W)}  \nn
&~~+ \frac{(\delta m^{(1)}_W)^2\, m_W\,\Gamma^0_W\, g^0_{W, f_L}\,g^0_{W, \ell_L}((\Gamma^0_W)^2(3\,\hat s^2 + 4\, m^2_W\, \hat s - 3\, m^4_W) - 6(m^2_W -\hat s)^2(m^2_W + \hat s)  - m^2_W\, (\Gamma^0_W)^4)}{P^3_0(\hat s, m_W, \Gamma^{(0)}_W)}  \nn
& ~~+ \frac{\delta m^{(2)}_W\,\Gamma^0_W\, g^0_{W, f_L}\, g^0_{W, \ell_L}\, (3m^4_W + m^2_W ((\Gamma^0_W)^2\, -2\,\hat s) - \hat s^2)}{P^2_0(\hat s, m_W, \Gamma^{(0)}_W)} \nn 
& ~~ + \frac{\delta m^{(1)}_W\,\delta \Gamma^{(1)}_W\,g^0_{W, f_L}\, g^0_{W, \ell_L} ((m^2_W - \hat s)^3\,(3\, m^2_W + \hat s) - m^4_W\, (\Gamma^0_W)^4 + 6\, m^2_W\, (\Gamma^0_W)^2\, (\hat s^2 - m^4_W))}{P^3_0(\hat s, m_W, \Gamma^{(0)}_W)}
\end{align}

\section{Results for Dilepton production $pp \to \ell^+\ell^-$}
\label{sec:dilepton}

The coupling expansions for $g_{Z, \psi_\lambda}$ for both electroweak input schemes have been presented in Ref.~\cite{Hays:2020scx} so we do not repeat them here. The $Z$ width (in the limit of massless fermions) can be found in Ref.~\cite{Corbett:2021eux}. Many of the results in the next few appendices can be adapted to dilepton production at $\ell^+\ell^-$ colliders; see Appendix~\ref{app:fourfermilepton} for details.

We have used a slightly different basis for the dimension eight operators compared to Ref.~\cite{2020,Boughezal:2021tih}. The majority of the differences are in the four fermion operators with two derivatives. Ref.~\cite{Boughezal:2021tih} use the combinations
$D_\mu(\psi^\dag \bar\sigma^\nu \psi)D^\mu(\chi^\dag \bar\sigma_\nu \chi)$ and $(\psi^\dag \bar\sigma^{(\nu}\overleftrightarrow{D}^{\mu)} \psi)(\chi^\dag \bar\sigma_{(\nu}\overleftrightarrow{D}_{\mu)} \chi)$, where $\gamma^{(\mu} \overleftrightarrow{D}^{\nu)} =\left(\gamma^{\mu} \overleftrightarrow{D}^{\nu}+\gamma^{\nu} \overleftrightarrow{D}^{\mu}\right)$, while we have used the set in Eq.~\eqref{eq:contactH3} to more easily track the $\hat s$ vs. $\hat t$ dependence. The two sets are easily related. The other difference is Ref.~\cite{2020,Boughezal:2021tih} keep operators of the form $D^3\psi^2H^2$, while we work in a basis where both of the Higgses carry derivatives in all $D^3\psi^2H^2$  operators. Explicitly,
\begin{align}
(\psi^{\dagger} \bar{\sigma}^{\mu} \psi)(D_{(\mu,} D_{\nu)} H^{\dagger})(D_{\nu} H) + h.c. \,\, \text{for\,} \psi = Q,L,u_c,d_c,e_c\nonumber \\
 (\psi^{\dagger} \bar{\sigma}^{\mu} D_{\nu} \psi)(D_{\mu} H^{\dagger})(D_{\nu} H) + h.c. \,\, \text{for\,} \psi = Q,L,u_c,d_c,e_c\nonumber \\
 (\psi^{\dagger} \bar{\sigma}^{\mu}\,\tau^I \psi)(D_{(\mu,} D_{\nu)} H^{\dagger})\tau_I\,(D_{\nu} H) + h.c.  \,\, \text{for\,} \psi = Q,L\nonumber \\
 (\psi^{\dagger} \bar{\sigma}^{\mu}\,\tau^I D_{\nu} \psi)(D_{\mu} H^{\dagger})\tau_I\,(D_{\nu} H) + h.c.  \,\, \text{for\,} \psi = Q,L,   
 \label{eq:d3geobasis}
\end{align}
where $D_{(\mu,} D_{\nu)}$ indicates symmetrized derivatives. In our basis, the $D^3\psi^2H^2$ operators all come with two gauge bosons, two Higgses, or one of each and cannot contribute to $pp \to \ell^+\ell^-$. The fact that one can always place the derivatives in this fashion (exploiting EOM and IBP) is guaranteed by the results of Ref.~\cite{Helset:2020yio}.  

Given a UV model that generates the $D^3\psi^2H^2$ operators used in Ref.~\cite{2020,Boughezal:2021tih}, manipulating these terms into a geoSMEFT compliant form via EOM and IBP will change the matching onto {\em other} operators (i.e. $D\psi^2H^4$ type). As a result, calculations of physical quantities in the geoSMEFT compliant basis will agree with results carried out in the original (Ref.~\cite{2020,Boughezal:2021tih}) basis. However, in the bottom up approach used in this paper, we ignore all relations among coefficients, so it appears as if the geoSMEFT compliant results depend on fewer operators than results in other bases\footnote{Note that Ref.~\cite{Hays:2020scx} also calculates dimension eight effects ( in the process $pp \to W h$) in a non geoSMEFT-compliant basis, so that $D^3\psi^2H^2$ effects feed into $ffV$ vertices. Applying the logic of the previous paragraphs to that example, had one calculated $pp \to W h$ using the basis of Eq.~\eqref{eq:d3geobasis}, Eq.~(4.1) of Ref.~\cite{Hays:2020scx} would not contain any dependence on $c_{qqV1}$. }.

Note that Ref.~\cite{2020,Boughezal:2021tih} included SM QCD loops in their calculation, while all of our results are leading order.

\subsection{Helicity amplitude expansion for $pp \to \ell^+\ell^-$}
\label{sec:dyhelas}
In the $\hat m_W$ scheme the helicity amplitudes are:
\label{sec:helampNC}
\begin{align}
& \text{Re}(\mathcal A^{(0)}_{\lambda_f,\lambda_\ell}) = \Big(\frac{(e^0)^2\,Q_f\,Q_\ell}{\hat s} + \frac{(\hat s - m^2_Z)\, g^0_{Z, f_{\lambda_f}}\, g^0_{Z, \ell_{\lambda_\ell}}}{P_0(\hat s, m_Z, \Gamma^0_Z)} \Big), \quad \text{Im}(\mathcal A^{(0)}_{\lambda_f,\lambda_\ell}) = -\frac{\Gamma^0_Z\, m_Z\, g^0_{Z, f_{\lambda_f}}\, g^0_{Z, \ell_{\lambda_\ell}} }{P_0(\hat s, m_Z, \Gamma^0_Z)} 
\end{align}
where $P_0(\hat s, m_Z, \Gamma^0_Z) = (m^2_Z - \hat s)^2 + m^2_Z\, (\Gamma^0_Z)^2$ and $\lambda_i, \lambda_F = L,R$ are the initial and final state helicities. All couplings and coupling shifts are real, which simplifies the expressions compared to the charged current case. At $\mathcal O(x)$:
\begin{align}
&\text{Re}(\mathcal A^{(1)}_{\lambda_f,\lambda_\ell}) = \Big( \delta e^{(1)} \frac{2\, e^0\, Q_f\, Q_\lambda}{\hat s} - \frac{(m^2_Z - \hat s)(\delta g^{(1)}_{Z, f_{\lambda_f}} g^0_{Z, \ell_{\lambda_\ell}} + \delta g^{(1)}_{Z, \ell_{\lambda_\ell}} g^0_{Z, f_{\lambda_f}}) }{P_0(\hat s, m_Z, \Gamma^0_Z)}  + \nn
&~~~~~~~~~~~~~~~~~~~~~~~~~~~~~~~~~~~~~~~~~~~~~~~~~~~~~~~~ \frac{2\,\delta \Gamma^{(1)}_Z\, \Gamma^0_Z\, m^2_Z g^0_{Z,f_{\lambda_f}}\,g^0_{Z, \ell_{\lambda_\ell}} (m^2_Z - \hat s)}{P^2_0(\hat s, m_Z, \Gamma^0_Z)} + \frac{\delta a^{(1)}_{\lambda_f,\lambda_\ell}}{v^2_T} \Big) \nn
&\text{Im}(\mathcal A^{(1)}_{\lambda_f,\lambda_\ell}) = \Big(- \frac{\Gamma^0_Z\, m_Z\,(\delta g^{(1)}_{Z, f_{\lambda_f}} g^0_{Z, \ell_{\lambda_\ell}} + \delta g^{(1)}_{Z, \ell_{\lambda_\ell}} g^0_{Z, f_{\lambda_f}})  }{P_0(\hat s, m_Z, \Gamma^0_Z)} -\frac{\delta \Gamma^{(1)}_Z\, m_Z\, g^0_{Z, f_{\lambda_f}} g^0_{Z, \ell_{\lambda_\ell}} ((m^2_Z - \hat s)^2 - m^2_Z\, (\Gamma^0_Z)^2)}{P^2_0(\hat s, m_Z, \Gamma^0_Z )} \Big) 
\end{align}
Finally, for $\mathcal O(x^2)$:
\begin{align}
&\text{Re}(\mathcal A^{(2)}_{\lambda_f,\lambda_\ell}) = \Big( (\delta e^{(1)})^2\frac{Q_f\,Q_\lambda}{\hat s} + 2\delta e^{(2)}\, \frac{e^0\, Q_f\, Q_\ell}{\hat s} - \frac{(m^2_Z - \hat s)(\delta g^{(1)}_{Z, f_{\lambda_f}} \delta g^{(1)}_{Z, \ell_{\lambda_\ell}} + \delta g^{(2)}_{Z, f_{\lambda_f}} g^0_{Z, \ell_{\lambda_\ell}} + \delta g^{(2)}_{Z, \ell_{\lambda_\ell}} g^0_{Z, f_{\lambda_f}} )}{P_0(\hat s, m_Z, \Gamma^0_Z)} + \nn
& ~~~~~~~~~~~~~~~~~~~~~\frac{2\delta \Gamma^{(1)}_Z\, \Gamma^0_Z\, m^2_Z\, (m^2_Z - \hat s)(\delta g^{(1)}_{Z, f_{\lambda_f}} g^0_{Z, \ell_{\lambda_\ell}} + \delta g^{(1)}_{Z, \ell_{\lambda_\ell}} g^0_{Z, f_{\lambda_f}})}{P^2_0(\hat s, m_Z, \Gamma^0_Z)} + \frac{2\,\delta \Gamma^{(2)}_Z\, \Gamma^0_Z\, m^2_Z\, (m^2_Z - \hat s) g^0_{Z, f_{\lambda_f}}g^0_{Z, \ell_{\lambda_\ell}}}{P^2_0(\hat s, m_Z, \Gamma^0_Z)} + \nn
& ~~~~~~~~~~~~~~~~~~~~~~\frac{(\delta\Gamma^{(1)}_Z)^2\, g^0_{Z, f_{\lambda_f}}g^0_{Z, \ell_{\lambda_\ell}}\, m^2_Z\, (m^2_Z - \hat s)(m^4_Z + \hat s^2 - m^2_Z(3(\Gamma^0_Z)^2 + 2\hat s) )}{P^3_0(\hat s, m_Z, \Gamma^0_Z)} + \nn
& ~~~~~~~~~~~~~~~~~~~~~~~~\frac 1 {\bar v^2_T}\Big( \delta a^{(2)}_{\lambda_f,\lambda_\ell} - \frac{\hat s }{\bar v^2_T}\delta a^{(2,s)}_{\lambda_f,\lambda_\ell} - \frac{\hat t }{\bar v^2_T}\delta a^{(2,t)}_{\lambda_f,\lambda_\ell}\Big) \\
&\text{Im}(\mathcal A^{(2)}_{\lambda_f,\lambda_\ell}) =  \Big( -\frac{\delta \Gamma^{(2)}_Z\, m_Z\, g^0_{Z, f_{\lambda_f}} g^0_{Z, \ell_{\lambda_\ell}} ((m^2_Z - \hat s)^2 - m^2_Z\, (\Gamma^0_Z)^2)}{P^2_0(\hat s, m_Z, \Gamma^0_Z )} - \nn
& ~~~~~~~~~~~~~~~~~~~~~~~~~~~~~~~~ \frac{\Gamma^0_Z\,m_Z\,(\delta g^{(1)}_{Z, f_{\lambda_f}} \delta g^{(1)}_{Z, \ell_{\lambda_\ell}} + \delta g^{(2)}_{Z, f_{\lambda_f}} g^0_{Z, \ell_{\lambda_\ell}} + \delta g^{(2)}_{Z, \ell_{\lambda_\ell}} g^0_{Z, f_{\lambda_f}} )}{P_0(\hat s, m_Z, \Gamma^0_Z)} - \nn
& ~~~~~~~~~~~~~~~~~~~~~~~~~~~~~ \frac{\delta\Gamma^{(1)}_Z\, m_Z\, (\delta g^{(1)}_{Z, f_{\lambda_f}} g^0_{Z, \ell_{\lambda_\ell}} + \delta g^{(1)}_{Z, \ell_{\lambda_\ell}} g^0_{Z, f_{\lambda_f}}) (m^4_Z + \hat s^2 - m^2_Z((\Gamma^0_Z)^2 + 2\hat s) )}{P^2_0(\hat s, m_Z, \Gamma^0_Z)}  + \nn
&~~~~~~~~~~~~~~~~~~~~~~~~~~~~~ \frac{(\delta \Gamma^{(1)}_Z)^2\,\Gamma^0_Z\, m^3_Z(3(m^2_Z - \hat s)^2 - (\Gamma^0_Z)^2\, m^2_Z) g^0_{Z, f_{\lambda_f}} g^0_{Z, \ell_{\lambda_\ell}} }{P^3_0(\hat s, m_Z, \Gamma^0_Z)} \Big)
\end{align}

The helicity amplitudes in the $\hat{\alpha}_{ew}$ scheme can be obtained from the expressions above by setting $\delta e^{(1)} = \delta e^{(2)} = 0$, as it is an input parameter. As $m_W$ does not appear, expanding it as a series in $x$ has no effect.

In either scheme, the pieces can be combined to form the amplitude squared to $\mathcal O(x^2)$:
\begin{align}
& |\mathcal A_{\lambda_f, \lambda_\ell}|^2 = (\text{Re}(\mathcal A^{(0)}_{\lambda_f, \lambda_\ell})^2 + \text{Im}(\mathcal A^{(0)}_{\lambda_f, \lambda_\ell})^2 ) + 2x\, (\text{Re}(\mathcal A^{(0)}_{\lambda_f, \lambda_\ell})\text{Re}(\mathcal A^{(1)}_{\lambda_f, \lambda_\ell}) + \text{Im}(\mathcal A^{(0)}_{\lambda_f, \lambda_\ell})\text{Im}(\mathcal A^{(1)}_{\lambda_f, \lambda_\ell}) \nn
&  ~~~~~ +  x^2\, \Big(2\, \text{Re}(\mathcal A^{(0)}_{\lambda_f, \lambda_\ell})\text{Re}(\mathcal A^{(2)}_{\lambda_f, \lambda_\ell}) + 2\, \text{Im}(\mathcal A^{(0)}_{\lambda_f, \lambda_\ell})\text{Im}(\mathcal A^{(2)}_{\lambda_f, \lambda_\ell}) + \text{Re}(\mathcal A^{(1)}_{\lambda_f, \lambda_\ell})^2 + \text{Im}(\mathcal A^{(1)}_{\lambda_f, \lambda_\ell})^2 \Big) 
\label{eq:ampsquared}
\end{align}

\subsection{Parton level calculation for $q \bar{q} \to \ell^+\ell^-$}
\label{sec:dileptonexplicitpartonlevel}

Plugging in to the expressions for the amplitude squared using the expressions for the contact terms (Sec.~\ref{sec:helampexpand}), couplings and $Z$ width, we find the partonic cross section. As in Sec.~\ref{sec:Wpparton}, we will examine the pieces that grow with energy. First, at $\mathcal O(x)$ there is a constant term, implying an amplitude that grows linearly with $\sqrt{\hat s}$,
\begin{align}
&\hat\sigma_{s\to\infty}(\bar u_i u_i  \to \ell^+\ell^-)_{\mathcal O(\hat s^0)} =  \frac{x\, \hat e^2}{432\,\pi\, \hat v^2\, \hat c^2_\theta}\Big( \frac{2 + \hat c_{2\theta}} {2\, \hat s^2_\theta} (C^{3,(6)}_{LQ} - C^{1,(6)}_{LQ} )-C^{(6)}_{eQ} - 4\, C^{(6)}_{eu} - 2\, C^{(6)}_{Lu} \Big)  + \mathcal O(x^2) \nn
&\hat\sigma_{s\to\infty}(\bar d_i d_i  \to \ell^+\ell^-)_{\mathcal O(\hat s^0)} =  \frac{x\, \hat e^2}{432\,\pi\, \hat v^2\, \hat c^2_\theta}\Big( \frac{1 + 2\hat c_{2\theta}} {2\, \hat s^2_\theta} (C^{3,(6)}_{LQ} + C^{1,(6)}_{LQ} )-C^{(6)}_{eQ} +2 \, C^{(6)}_{ed} + C^{(6)}_{Ld} \Big)  + \mathcal O(x^2) 
\label{eq:DYos0}
\end{align}
where $u_i (d_i)$ stand for up and charm quarks (down and strange). As opposed to monolepton production, there are multiple operators present and with different signs making accidental cancellations a possibility. Judiciously choosing signs of coefficients, it is possible to make all of the operators in $\hat\sigma(\bar u_i u_i  \to \ell^+\ell^-)$ {\em or} $\hat\sigma_{s\to\infty}(\bar d_i d_i  \to \ell^+\ell^-)$ positive, but not both. 

Moving to the $\mathcal O(\hat s)$ cross section terms -- meaning an amplitude that grows as $\hat s$ -- which arise at $\mathcal O(x^2)$:
\begin{align}
&\hat\sigma_{s\to\infty}(\bar u_i u_i  \to \ell^+\ell^-)_{\mathcal O(\hat s)} = \frac{x^2\,\hat s}{3456\, \pi\, \hat v^4\, \hat c^2_\theta}\Big(24\,\hat c^2_{\theta}\,\Big(C^{(6)}_{eQ})^2 + (C^{(6)}_{eu})^2 + (C^{(6)}_{Lu})^2 + (C^{3,(6)}_{LQ} - C^{1,(6)}_{LQ})^2\Big)\nn
& ~~~~~~~~~~~~~~~~~~~ + \hat e^2\Big(8\, C^{s,(8)}_{eQ} - 6\, C^{t,(8)}_{eQ} + 32\, C^{s, (8)}_{eu} - 8\, C^{t,(8)}_{eu} + 16\, C^{s,(8)}_{Lu} - 12\, C^{t,(8)}_{Lu} \nn
& ~~~~~~~~~~~~~~~~~~~~~~~~ +  (C^{s,3,(8)}_{LQ} - C^{s,1,(8)}_{LQ})(8 - 12/\hat s^2_\theta) + (C^{t,1,(8)}_{LQ} - C^{t,3,(8)}_{LQ})(2 - 3/\hat s^2_\theta)  \Big)\Big) \nn
&\hat\sigma_{s\to\infty}(\bar d_i d_i  \to \ell^+\ell^-)_{\mathcal O(\hat s)} = \frac{x^2\,\hat s}{3456\, \pi\, \hat v^4\, \hat c^2_\theta}\Big(24\,\hat c^2_{\theta}\,  \Big(C^{(6)}_{eQ})^2 + (C^{(6)}_{ed})^2 + (C^{(6)}_{Ld})^2 + (C^{3,(6)}_{LQ} + C^{1,(6)}
_{LQ})^2\Big)\nn
& ~~~~~~~~~~~~~~~~~~~ + \hat e^2\Big(8\, C^{s,(8)}_{eQ} - 6\, C^{t,(8)}_{eQ} - 16\, C^{s, (8)}_{ed} + 4\, C^{t,(8)}_{ed} - 8\, C^{s,(8)}_{Ld} + 6\, C^{t,(8)}_{Ld} \nn
& ~~~~~~~~~~~~~~~~~~~~~~~~ +  (C^{s,3,(8)}_{LQ} + C^{s,1,(8)}_{LQ})(16 - 12/\hat s^2_\theta) + (C^{t,1,(8)}_{LQ} + C^{t,3,(8)}_{LQ})(3/\hat s^2_\theta - 4)  \Big)\Big)
\label{eq:DYox2}
\end{align}

\subsection{Proton level $pp \to \ell^+\ell^-$ results}
\label{sec:DYfigs}

In this section we show some illustrative proton level results for $pp \to \ell^+\ell^-$. We make the same parton distribution choices as in Sec.~\ref{sec:results}. 

\begin{figure}[h!]
\includegraphics[width=0.49\textwidth]{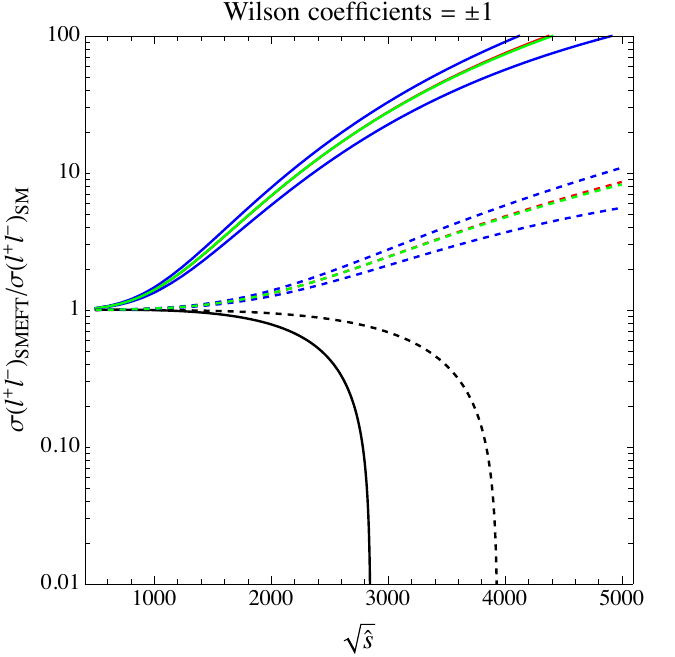}
\includegraphics[width=0.49\textwidth]{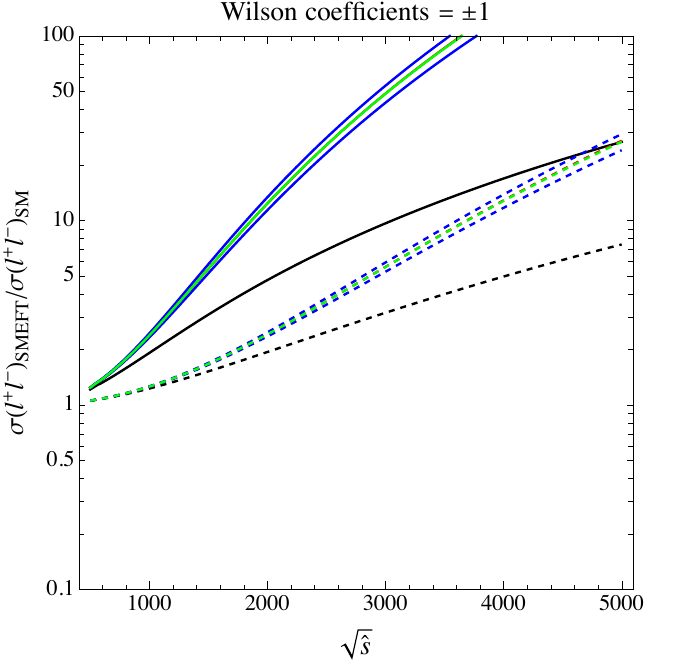}
\caption{$d\sigma(pp \to \ell^+\ell^-)/d\sqrt{\hat s}$ at various SMEFT orders compared to the SM assuming all Wilson coefficients have magnitude $1$. The black line shows the result including $\mathcal O(1/\Lambda^2)$ SMEFT terms while the green and blue lines show the ratios including dimension six squared terms and the full  $\mathcal O(1/\Lambda^4)$ calculation, respectively. The two $\mathcal O(1/\Lambda^4)$ curves differ in the assumptions made about the sign of the dimension eight pieces; in the upper curve, we chose signs to maximize Eq.~\eqref{eq:DYox2} (as much as possible, choosing to maximize up quark initial states over down quark), while in the lower curve we minimize Eq.~\eqref{eq:DYox2}.  The curve with all all dimension eight coefficients chosen positive is nearly identical to the green line, as it is subject to cancellations. For the solid lines, the new physics scale $\Lambda = 5\, \text{TeV}$, while dashed lines have $\Lambda = 10\, \text{TeV}$. The difference between the left and right panels is the sign choices for the dimension six contact terms. In the left panel, all coefficients are taken as $+1$, which causes cancellations within some helicity pieces and results in a net destructive interference with the SM. In the right panel, the signs have been chosen to make the interference (at least for up quark initial states) positive definite.
}
\label{fig:DYdshat}
 \end{figure}

In Fig.~\ref{fig:DYdshat} below we show the differential cross section as a function of the dilepton invariant mass. As the large $\hat s$ pieces of the partonic amplitudes involve multiple coefficients, the differential cross section is more susceptible to coefficient sign choices than $pp \to \ell^+\nu^-$. To illustrate this, we show the differential cross section for two different sign choices; in the left panel, all dimension six coefficients have been take to be $+1$, while in the right panel we have chosen the sign convention to maximize the number of positive $\mathcal O(s)$ terms. 
When all dimension six coefficients are positive, there is a net destructive interference. Once we include the positive definite dimension six squared piece at $\mathcal O(1/\Lambda^4)$, the SMEFT result is again greater than the SM, but the shift from destructive to constructive means an exaggerated ration when we compare the two, $(\sigma(pp \to \ell^+\ell^-)_{\mathcal O(1/\Lambda^4)} - \sigma(pp \to \ell^+\ell^-)_{\mathcal O(1/\Lambda^2)})/\sigma(pp \to \ell^+\ell^-)_{\mathcal O(1/\Lambda^2)}$. If we choose the dimension six coefficients such that the $\mathcal O(1/\Lambda^2)$ is positive, the full $\mathcal O(1/\Lambda^4)$ result is similar to the previous choice, but the ratio of $\mathcal O(1/\Lambda^4)$ to $\mathcal O(1/\Lambda^2)$ results is much smaller. We see this trend in Fig.~\ref{fig:DYratiosLam4} below, where we plot $(\sigma(pp \to \ell^+\ell^-)_{\mathcal O(1/\Lambda^4)}-\sigma(pp \to \ell^+\ell^-)_{\mathcal O(1/\Lambda^2)})/\sigma(pp \to \ell^+\ell^-)_{\mathcal O(1/\Lambda^2)}$ as a function of the minimum and maximum dilepton invariant mass.

\begin{figure}[h!]
\includegraphics[width=0.49\textwidth]{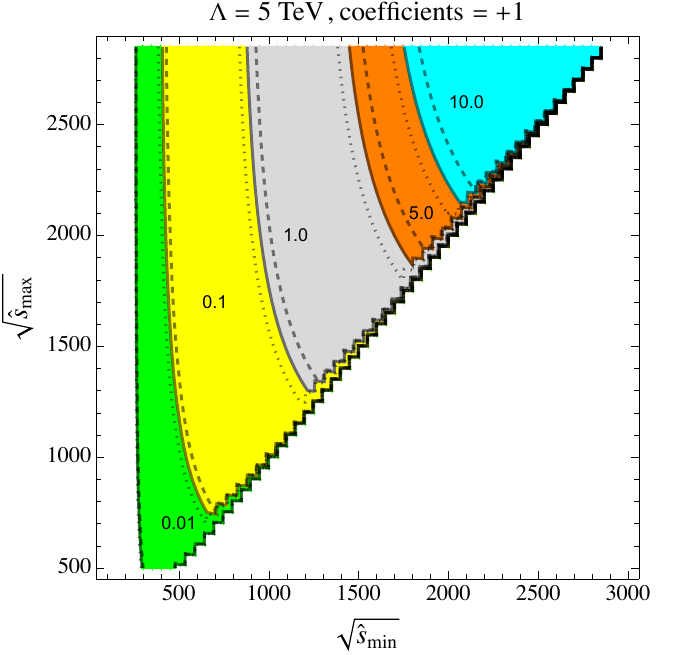}
\includegraphics[width=0.49\textwidth]{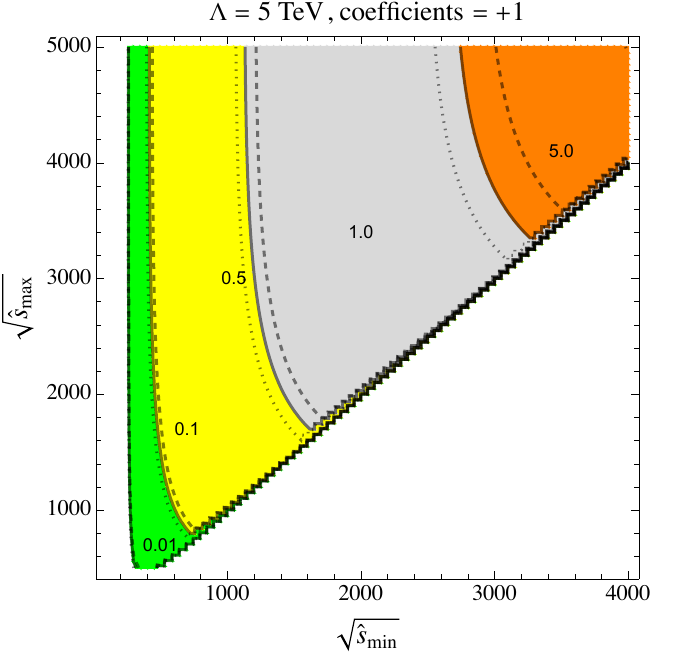}
\caption{Ratio of the $pp \to \ell^+\ell^-$ cross section at $\mathcal O(1/\Lambda^4)$ to the cross section at $\mathcal O(1/\Lambda^2)$ as a function of the minimum and maximum $\sqrt{\hat s}$. In the left panel, all dimension six contact terms coefficients are $+1$, while in the right hand plot the sign of the coefficient has been chosen following Eq.~\eqref{eq:DYos0} to give positive interference. The dashed and dotted lines are the contours after choosing dimension eight coefficient signs to maximize (dotted) or minimize (dashed) Eq.~\ref{eq:DYox2}. For both panels the new physics scale $\Lambda = 5\, \text{TeV}$. The axes range in the left panel is smaller because the $\mathcal O(1/\Lambda^2)$ cross section for that coefficient sign choice becomes negative above $\sqrt{\hat s} \sim 3\, \text{TeV}$.}
\label{fig:DYratiosLam4}
\end{figure}

Figure~\ref{fig:DYratiosLam4} is the $pp \to \ell^+\ell^-$ analog of the left panel of Fig.~\ref{fig:ratiosWp}. As in Fig.~\ref{fig:DYdshat}, the two panels in Fig.~\ref{fig:DYratiosLam4} correspond to two different choices for the signs of the dimension six contact terms. Comparing this figure with the monolepton counterpart (Fig.~\ref{fig:DYratiosLam4}), see that $\mathcal O(1/\Lambda^4)$ terms generally have a larger effect (under our coefficient assumptions), even when the dimension six coefficient signs have been chosen to make the interference positive. In the left panel, the $\sqrt{\hat s_{min}}$ and $\sqrt{\hat s_{max}}$ have been restricted to a range where the $\mathcal O(1/\Lambda^2)$ cross section is positive.

\begin{figure}[h!]
\includegraphics[width=0.49\textwidth]{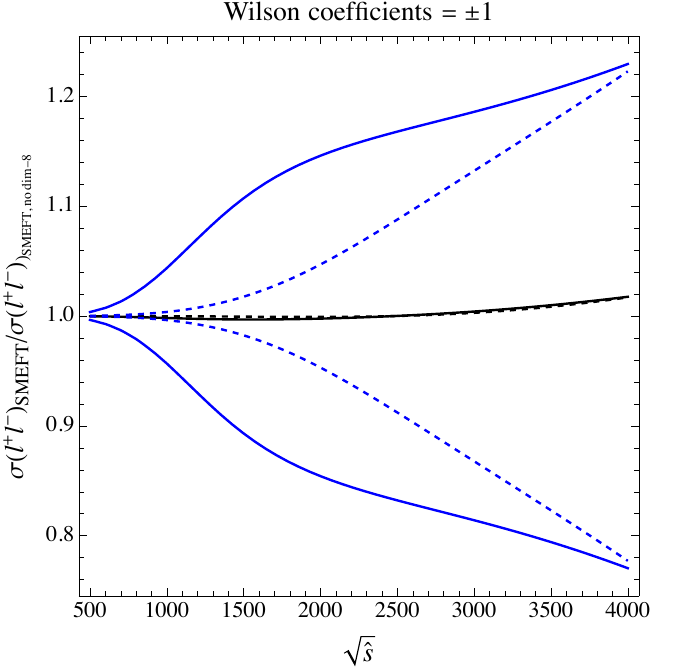}
\includegraphics[width=0.49\textwidth]{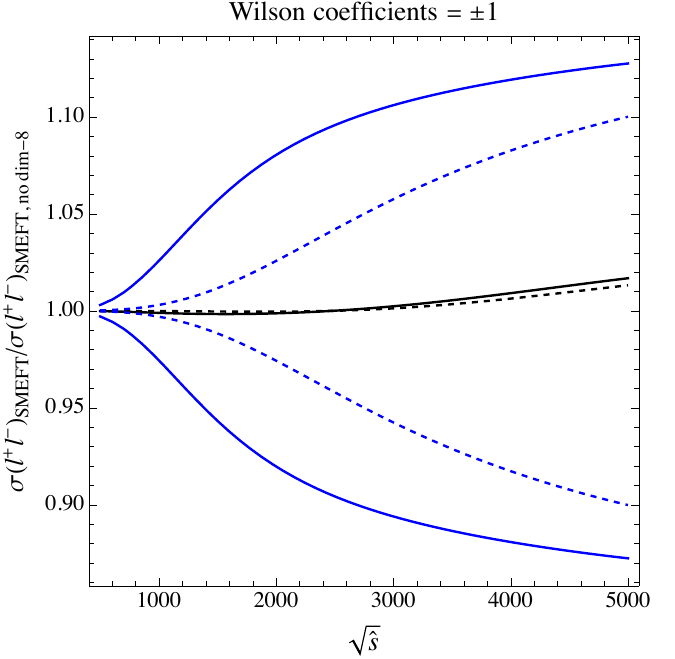}
\caption{Dimension eight contributions to $pp \to \ell^+\ell^-$ as a function of $\sqrt{\hat s}$ relative to the $\mathcal O(1/\Lambda^4)$ result from dimension six operators alone. In the black line we show the ratio with all coefficients set to $+1$ while in the blue lines show the ratio when the coefficient signs have been chosen to maximize (upper curve) or minimize (lower curve) Eq.~\eqref{eq:DYox2}. The solid lines assume a new physics scale of $\Lambda = 5\,\text{TeV}$, while the dashed line assumes $\Lambda = 10\, \text{TeV}$. The difference between the left and right panels is the assumption about the dimension six coefficient signs, as in Fig.~\ref{fig:DYdshat}.}
\label{fig:DYratioPS}
\end{figure}

Next, to zoom in on the role of the dimension eight operators, in Fig.~\ref{fig:DYratioPS} we have plotted the ratio of the full $\mathcal O(1/\Lambda^4)$ result to the $\mathcal O(1/\Lambda^4)$ using only dimension six squared. As with Fig.~\ref{fig:DYdshat}, we show results for two different new physics scales and for two different choices for the signs of the dimension six contact terms. 
For both dimension six sign choices we see that setting all dimension eight coefficients to $+1$ leads to cancellations among those terms and a suppressed net dimension eight effect. Flipping signs so that as much of the dimension eight contribution is positive (upper curve) or negative (lower curve) breaks the cancellations, and the importance of the dimension eight rises with center of mass energy. When the dimension six contribution is subject to cancellation (left panel), the dimension eight piece is more significant, reaching $\mathcal O(10\%)$ for $\sqrt{\hat s} = 4\, \text{TeV}$ for both $\Lambda$ scales we've considered; when the dimension six interference is positive, the dimension eight effects reach only $\mathcal O(6\%)$ for $\sqrt{\hat s} = 4\, \text{TeV}$ for $\Lambda = 5\, \text{TeV}$,  smaller for larger $\Lambda$. 

Finally, we can estimate the effects of $\mathcal O(1/\Lambda^6)$ using the procedure at the end of Sec.~\ref{sec:allorders}. All four helicity combinations contribute, and the terms are different for up-type and down-type initial states. Making the approximation that only a single electroweak structure enters at dimension ten (to avoid cancellations) and taking all Wilson coefficients to be +1:
\begin{align}
\hat\sigma(\bar u u \to \ell^+\ell^-)_{\mathcal O(x^3), C_i = 1} &= \frac{ s^2\, x^3}{5760\, \pi\, \hat v^6}\Big(60+17\, \frac{\hat e^2}{\hat c^2_{\theta}\hat s^2_{\theta}} (2\,\hat c_{2\theta}-3) \Big) \\
\hat\sigma(\bar d d \to \ell^+\ell^-)_{\mathcal O(x^3), C_i = 1} &= \frac{s^2\, x^3}{5760\, \pi\, \hat v^6}\Big(-60 + 17\,\frac{\hat e^2}{\hat c^2_{\theta}\hat s^2_{\theta}} \Big)
\end{align}

\begin{figure}[h!]
\includegraphics[width=0.49\textwidth]{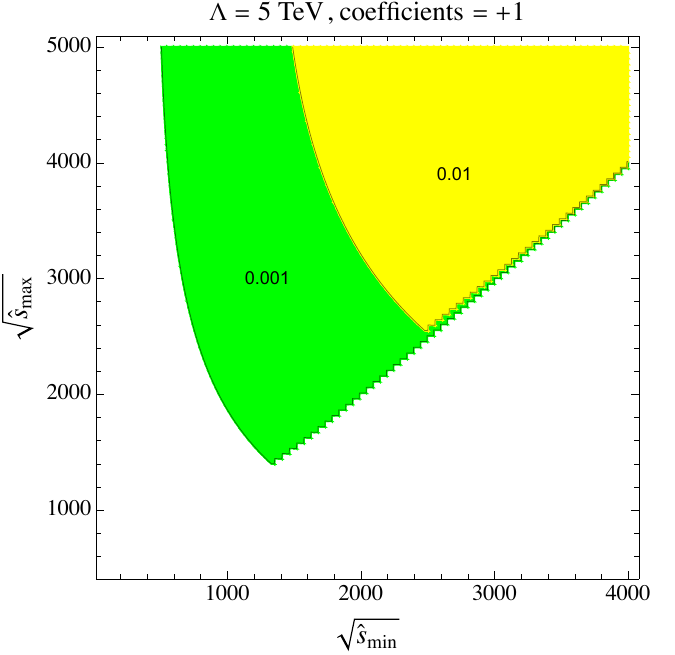}
\caption{Ratio of the $\mathcal O(1/\Lambda^6)$ contributions relative to the SMEFT dilepton cross section to $\mathcal O(1/\Lambda^4)$ as a function of the minimum center of mass energy $\sqrt{\hat s_{min}}$ and the maximum $\sqrt{\hat s_{max}}$, $(\sigma(pp \to \ell^+\ell^-)_{\mathcal O(1/\Lambda^6)} - \sigma(pp \to \ell^+\ell^-)_{\mathcal O(1/\Lambda^4)})/\sigma(pp \to \ell^+\ell^-)_{\mathcal O(1/\Lambda^4)}$. All Wilson coefficients are taken to be positive and equal to one, and the new physics scale $\Lambda = 5\,\text{TeV}$.} 
\label{fig:ratiodim10DY}
\end{figure}

Using these results, in Fig.~\ref{fig:ratiodim10DY}, we quantify the $\mathcal O(1/\Lambda^6)$ effects by taking the ratio $|(\sigma(pp \to \ell^+ \ell^-)_{\mathcal O(x^3)} - \sigma(pp \to \ell^+ \ell^-)_{\mathcal O(x^2)})/\sigma(pp \to \ell^+\ell^-)_{\mathcal O(x^2)}|$ as a function of the minimum and maximum center of mass energy (the dilepton version of Fig.~\ref{fig:ratiodim10} and taking $\Lambda = 5\, \text{TeV}$. 

As in Fig.~\ref{fig:ratiodim10}, we have approximated the full $\mathcal O(1/\Lambda^6)$ result with the dominant piece in the large $\hat s$ limit. For this simple Wilson coefficient choice, we find that the $\mathcal O(1/\Lambda^6)$ are significantly smaller than the $\mathcal O(1/\Lambda^4)$ for the kinematic region we have explored.

\section{Contact operators for $\ell^+_i \ell^-_i \to \ell^+_j \ell^-_j$ at dimension six and eight}
\label{app:fourfermilepton}

The neutral current results of this paper can easily be extended to lepton colliders (under the same flavor assumptions), $\ell^+ \ell^- \to \bar q_i\ q_i$ and $\ell_i^+ \ell_i^- \to \ell^+_j\ \ell^+_j$ where $i,j$ are flavor labels. For $\ell^+ \ell^- \to \bar q_i\ q_i$, the partonic amplitudes are identical to Sec.~\ref{sec:dilepton}, as all we have to do is switch the initial and final states. For $\ell^+_i \ell^-_i \to \ell^+_j\ \ell^-_j$, we need to swap $g^{0,(1),(2)}_{Z, f_\lambda} \to g^{0,(1),(2)}_{Z, \ell_\lambda} $ in the coupling expressions and to replace the two-quark, two-lepton contact terms with four lepton contact terms. 

When the initial and final lepton flavors are different, e.g. $e^+e^- \to \mu^+\mu^-$ the counting is similar to the two-quark, two-lepton case, but the exact number depends on what flavor assumptions we make. If we assume individual lepton number is conserved $U(1)_e \times U(1)_\mu \times U(1)_\tau$, the dimension six four fermion operators that contribute to $e^+e^- \to \mu^+\mu^-$ are:
\begin{align}
& C^{1,(6)}_{L_1L_2}\, ( L_1^\dag \,\bar\sigma^\mu\,  L_1)(L_2^\dag \,\bar\sigma_\mu\,  L_2) +  C^{3,(6)}_{L_1L_2}\, ( L_1^\dag \,\bar\sigma^\mu\, \tau^I L_1)(L_2^\dag \,\bar\sigma_\mu\, \tau_I L_2)\nn
& + C^{(6)}_{L_1e_2}\,  ( L_1^\dag \,\bar\sigma^\mu\,  L_1)(e^\dag_{c,2} \,\bar\sigma_\mu\, e_{c,2}) + C^{(6)}_{L_2e_1}\,  ( L_2^\dag \,\bar\sigma^\mu\,  L_2)(e^\dag_{c,1} \,\bar\sigma_\mu\, e_{c,1})  \nn
& + C^{(6)}_{L_1L_2e_1e_2} ( L_1^\dag \,\bar\sigma^\mu\,  L_2)(e^\dag_{c,2} \,\bar\sigma_\mu\, e_{c,1}) + h.c.  \nn
& + C^{(6)}_{e_1e_2}\, (e^\dag_{c,1} \,\bar\sigma_\mu\, e_{c,1})(e^\dag_{c,2} \,\bar\sigma_\mu\, e_{c,2}),
\label{eq:eemumucontacts6}
\end{align}
where we have explicitly plugged in generation indices to avoid confusion, and operators for other flavor combinations (i.e. $e^+e^-\to \tau^+\tau^-$) are identical except for the generation indices. Strengthening the flavor symmetry to $(U(1)_L)^3 \times (U(1)_e)^3$ removes the $C^{(6)}_{L_1L_2e_1e_2}$ operators, while enforcing $U(3)_L \times U(3)_e$ also sets $C^{(6)}_{L_1e_2} = C^{(6)}_{L_2e_1} $


At dimension eight and imposing $U(1)_e \times U(1)_\mu \times U(1)_\tau$, the  $e^+e^- \to \mu^+\mu^-$ operators containing two Higgses are:
\begin{align}
& C^{1,(8)}_{HL_1L_2}\, (H^\dag H)( L_1^\dag \,\bar\sigma^\mu\,  L_1)(L_2^\dag \,\bar\sigma_\mu\,  L_2)  + C^{2,(8)}_{HL_1L_2}\, (H^\dag H)( L_1^\dag \,\bar\sigma^\mu\, \tau^I L_1)(L_2^\dag \,\bar\sigma_\mu\, \tau_I L_2)\nn
&~~~~~~~~~~~~+  C^{3,(8)}_{HL_1L_2}\, (H^\dag \tau^I H)( L_1^\dag \,\bar\sigma^\mu\, \tau_I L_1)(L_2^\dag \,\bar\sigma_\mu\,  L_2)+ C^{4,(8)}_{HL_1L_2}\, (H^\dag \tau^I H)( L_1^\dag \,\bar\sigma^\mu\,  L_1)(L_2^\dag \,\bar\sigma_\mu\, \tau_I L_2)\nn
&~~~~~~~~~~~~~~~~~~~~~~~~~~~~~~+ C^{5,(8)}_{HL_1L_2}\, \epsilon_{IJK}  (H^\dag \tau^I H)( L_1^\dag \,\bar\sigma^\mu\, \tau^J  L_1)(L_2^\dag \,\bar\sigma_\mu\, \tau^K L_2)\nn
& ~~~~~~~~~~~ + C^{1,(8)}_{HL_1e_2}\, (H^\dag H)( L_1^\dag \,\bar\sigma^\mu\,  L_1)(e^\dag_{c,2} \,\bar\sigma_\mu\, e_{c,2}) +  C^{1,(8)}_{HL_2e_1}\, (H^\dag H)( L_2^\dag \,\bar\sigma^\mu\,  L_2)(e^\dag_{c,1} \,\bar\sigma_\mu\, e_{c,1}) \nn
& ~~~~~~ + C^{2,(8)}_{HL_1e_2}\, (H^\dag \tau^I H)( L_1^\dag \,\bar\sigma^\mu\, \tau_I L_1)(e^\dag_{c,2} \,\bar\sigma_\mu\, e_{c,2}) + C^{2,(8)}_{HL_1e_2}\, (H^\dag \tau^I H)( L_2^\dag \,\bar\sigma^\mu\, \tau_I L_2)(e^\dag_{c,1} \,\bar\sigma_\mu\, e_{c,1})  \nn
& ~~~  C^{1,(8)}_{L_1L_2e_1e_2} (H^\dag H)(L^{\dag}_1 \bar\sigma^\mu L_2)(e^{\dag}_{c,1}\bar\sigma_\mu e_2)  + C^{2,(8)}_{L_1L_2e_1e_2} (H^\dag \tau^I H)(L^{\dag}_1 \bar\sigma^\mu\,\tau^I L_2)(e^{\dag}_{c,1}\bar\sigma_\mu e_2)  + h.c. \nn
& ~~~~~~~~~~~~~~~~~~~~~~~~~~~~~~+ C^{(8)}_{He_1e_2}\,  (H^\dag H)(e^\dag_{c,1} \,\bar\sigma_\mu\, e_{c,1})(e^\dag_{c,2} \,\bar\sigma_\mu\, e_{c,2})\,,
\end{align}
while the operators with two derivatives are:
\begin{align}
& C^{s,1,(8)}_{L_1L_2}\, (L_1^{\dag}\, \bar{\sigma}^{\mu}\,  L_1)D^2(L_2^{\dag}\, \bar{\sigma}_{\mu}\, L_2) +  C^{t,1,(8)}_{L_1L_2}\, \Big((D_\nu L_1^{\dag}\, \bar{\sigma}^{\mu}\,  L_1)(D^{\nu}L_2^{\dag}\, \bar{\sigma}_{\mu}\, L_2) + h.c. \Big) \nn
& ~~~~+ C^{s,3,(8)}_{L_1L_2}\, (L_1^{\dag}\, \bar{\sigma}^{\mu}\,\tau^I  L_1)D^2(L_2^{\dag}\, \bar{\sigma}_{\mu}\,\tau_I L_2) + C^{t,1,(8)}_{L_1L_2}\, \Big((D_\nu L_1^{\dag}\, \bar{\sigma}^{\mu}\,  L_1)(D^{\nu}L_2^{\dag}\, \bar{\sigma}_{\mu}\, L_2) + h.c. \Big) \nn
& ~~~~~~+ C^{s,(8)}_{L_1e_2}\, ( L_1^{\dag}\, \bar{\sigma}^{\mu}\,  L_1)D^2(e^\dag_{c,2}\, {\bar\sigma}_{\mu}\, e_{c,2}) + C^{s,(8)}_{L_2e_1}\, ( L_2^{\dag}\, \bar{\sigma}^{\mu}\,  L_2)D^2(e^\dag_{c,1}\, {\bar\sigma}_{\mu}\, e_{c,1}) \nn
& ~~~+ C^{t,(8)}_{L_1e_2}\, \Big((D_\nu L_1^{\dag}\, \bar{\sigma}^{\mu}\,  L_1)(D^\nu e^\dag_{c,2}\, {\bar\sigma}_{\mu}\, e_{c,2}) + h.c. \Big) + C^{t,(8)}_{L_2e_1}\, \Big((D_\nu L_2^{\dag}\, \bar{\sigma}^{\mu}\,  L_2)(D^\nu e^\dag_{c,1}\, {\bar\sigma}_{\mu}\, e_{c,2}) + h.c. \Big) \nn
& + C^{s,(8)}_{L_1L_2e_1e_2} (L^{\dag}_1\bar{\sigma}^\mu L_2)D^2(e^\dag_{c,1}\bar\sigma_\mu e_{c,2}) + C^{t,(8)}_{L_1L_2e_1e_2} (D_\nu L^\dag_{1}\, {\bar\sigma}^{\mu}\,  L_{2})(D^{\nu} e^\dag_{c,1}\, {\bar\sigma}_{\mu}\, e_{c,2}) + h.c.   \nn
& +C^{s,(8)}_{e_1e_2}\, (e^\dag_{c,1}\, {\bar\sigma}^{\mu}\, e_{c,1})D^2(e^\dag_{c,2}\, {\bar\sigma}_{\mu}\, e_{c,2}) + C^{t,(8)}_{e_1e_2}\, \Big( (D_\nu e^\dag_{c,1}\, {\bar\sigma}^{\mu}\,  e_{c,1})(D^{\nu} e^\dag_{c,2}\, {\bar\sigma}_{\mu}\, e_{c,2}) + h.c. \Big)\,. 
\label{eq:eemumucontacts8}
\end{align}
In the above lists, all coefficients are real except for the $C^{i,(8)}_{L_1L_2e_1e_2}$. Under stronger flavor assumptions $(U(1)_L)^3 \times (U(1)_e)^3$ or $U(3)_L \times U(3)_e$, the number of dimension eight terms is reduced following the same pattern as at dimension six.

These four lepton operators translate into amplitude contact term coefficients (as in Eq.~\eqref{eq:expand2}) as:
\begin{align}
\delta a^{(1)}_{e_L\mu_L} =  C^{1,(6)}_{L_1L_2} + C^{3,(6)}_{L_1L_2},\quad \delta a^{(1)}_{e_L\mu_R} =  C^{(6)}_{L_1e_2},\quad \delta a^{(1)}_{e_R\mu_L} =  C^{(6)}_{L_2e_1},\quad \delta a^{(1)}_{e_R\mu_R} =  C^{(6)}_{e_1e_2}
\end{align}
\begin{align}
& \delta a^{(2)}_{e_L\mu_L} =  \frac{C^{1,(6)}_{L_1L_2} + C^{2,(6)}_{L_1L_2} + C^{3,(6)}_{L_1L_2} + C^{4,(6)}_{L_1L_2}}{2}, \quad \delta a^{(2,s)}_{e_L\mu_L} = C^{s,1,(8)}_{L_1L_2} + C^{s,3,(8)}_{L_1L_2}, \quad \delta a^{(2,t)}_{e_L\mu_L} = C^{t,1,(8)}_{L_1L_2} + C^{t,3,(8)}_{L_1L_2} \nonumber \\
&~~~~~~~~~~~~~~~~~~~~~~~~~~ \delta  a^{(2)}_{e_L\mu_R} = \frac{C^{1,(8)}_{L_1e_2} + C^{2,(8)}_{L_1e_2} }{2}, \quad  \delta  a^{(2,s)}_{e_L\mu_R} = C^{s,(8)}_{L_1e_2}, \quad  \delta  a^{(2,t)}_{e_L\mu_R} = C^{t,(8)}_{L_1e_2} \nonumber \\
& ~~~~~~~~~~~~~~~~~~~~~~~~~~ \delta  a^{(2)}_{e_R\mu_L} = \frac{C^{1,(8)}_{L_2e_1} + C^{2,(8)}_{L_2e_1} }{2}, \quad  \delta  a^{(2,s)}_{e_R\mu_L} = C^{s,(8)}_{L_2e_1}, \quad  \delta  a^{(2,t)}_{e_R\mu_L} = C^{t,(8)}_{L_2e_1} \nonumber \\
& ~~~~~~~~~~~~~~~~~~~~~~~~~~~~~~~ \delta  a^{(2)}_{e_R\mu_R} = \frac{C^{(8)}_{e_1e_2}  }{2}, \quad  \delta  a^{(2,s)}_{e_R\mu_R} = C^{s,(8)}_{e_1e_2}, \quad  \delta  a^{(2,t)}_{e_R\mu_R} = C^{t,(8)}_{e_1e_2}
\end{align}
Note the $C_{L_1L_2e_1e_2}$ do not appear because we have only listed the amplitudes with the same helicity structure as the SM. While dimension eight operators must have the same helicity structure ass the SM to contribute at $\mathcal O(1/\Lambda^4)$, dimension six operators do not. The flavor assumptions we made in the main text forbid the analogous terms from monolepton and dilepton production, but they are permitted under $U(1)_e \times U(1)_\mu \times U(1)_\tau$. In this case, the amplitude list from Appendix~\ref{sec:dyhelas} must be extended to include:
\begin{align}
A_{e_L\mu_R e_R \mu_L} = \frac{x}{\bar v^2_T} \delta a^{(1)}_{LRRL} S_{e_L\mu_R e_R \mu_L} ,\qquad \delta a^{(1)}_{LRRL} = C^{(6)}_{L_1L_2e_1e_2}\,.
\end{align}

If we take all four leptons to have the same flavor, e.g. for $e^+e^- \to e^+e^-$, Fermi statistics reduces the number of terms. Specifically, the number of $(H^\dag H) (L_i^\dag L_i)^2$ and  $D^2(L_i^\dag L_i)^2$ operators are both reduced to two, and the $L^\dag_i L_i e^\dag_{c,i}e_{c,i}$ operators collapse into a single form.

\bibliographystyle{JHEP}
\bibliography{bibliography.bib}

\end{document}